\documentclass[11pt]{article}  
\usepackage{graphicx}
\newcommand{\be}{\begin{equation}}
\newcommand{\ee}{\end{equation}}
\newcommand{\bea}{\begin{eqnarray}}
\newcommand{\eea}{\end{eqnarray}}

\begin{document}
\vspace{0.5in}
\vspace{.5in} 
\begin{center} 
{\LARGE{\bf Explicit Kink Solutions in Several One-Parameter Family of Higher Order Field Theory Models}}
\end{center} 

\vspace{0.2in} 

\begin{center}
{{\bf Avinash Khare}} \\
{Physics Department, Savitribai Phule Pune University \\
 Pune 411007, India}
\end{center}

\begin{center}
{{\bf Ayhan Duzgun and Avadh Saxena}} \\ 
{Theoretical Division and Center for Nonlinear Studies, 
Los Alamos National Laboratory, Los Alamos, New Mexico 87545, USA}
\end{center}

\vspace{0.2in}

\noindent{\bf {Abstract:}}
We present several one-parameter family of higher order field theory  
models some of which admit explicit kink solutions with an exponential tail while others admit explicit kink solutions with a power-law tail. Various properties of these families of kink solutions are examined in detail. In particular for models harboring kink solutions with a power-law tail, we show that there is no gap between the zero mode and the beginning of the continuum in the kink stability potential. Further, by applying the recent Manton formalism, we provide estimates for the kink-kink and antikink-kink acceleration and hence the ratio of the corresponding antikink-kink and kink-kink forces. 

\section{Introduction} 
During the last few years there has been a resurgence in studying kink solutions with exponential as well as power-law tails in several higher (than 6th) order field theory models including $\phi^8$, $\phi^{10}$, and $\phi^{12}$ \cite{KCS, Chapter}. The study of higher order field theories, their attendant kink excitations as well as the associated kink interactions and scattering \cite{scattering1, scattering2} are important in a variety of physical contexts ranging from successive phase transitions \cite{KCS, Chapter, Gufan82, Gufan78} to isostructural phase transitions \cite{Pavlov99} to models involving long-range interaction between massless mesons \cite{Lohe79}, as well as from protein crystallization \cite{Boulbitch97} to successive phase transitions presumably driving the late time expansion of the Universe \cite{Greenwood09}. Thus, understanding kink behavior in these models provides useful insight into the properties of domain walls in materials, condensed matter, high energy physics, biology and cosmology, which also serves as one of our main motivations here.

So far implicit kink solutions have been obtained in these higher order models using which one has studied in detail \cite{KS19} several aspects of these kink solutions such as the nature of kink tail, kink-kink and kink-antikink forces \cite{Manton19, Radomskiy17, ChristovPRD, ChristovPRL}, stability analysis, etc. However, one important missing piece has been the absence of analytically solvable explicit kink solutions in these higher order field theory models. Few exceptions include explicit kink solutions in a one-parameter family of $\phi^{4n+2} - \phi^{2n+2} - \phi^2$ models ($n = 1, 2, 3$) [1, 9], and one example of $\phi^8$ \cite{GaniPRD20} as well as one example of $\phi^{10}$ \cite{KS19} field theory models. The main objective of this paper is to partially fill this gap.

In this context it is worth pointing out that during the last four decades explicit kink solutions have been obtained in a number of lower order field theory models like $\phi^4$, $\phi^6$ \cite{Rajaraman} and also several non-polynomial hyperbolic \cite{hyperbolic} and periodic potentials. Beyond sine-Gordon \cite{Rajaraman}, the latter include double sine-Gordon \cite{Leung82} and Lam\'e \cite{periodic} solitons as examples. Further, recently explicit kink solutions with super-exponential \cite{PKS19, Flores02} and super-super-exponential \cite{KS20PS}, power-tower \cite{KS20JPA} as well as power-law tail \cite{Gomes12, Bazeia18, Guerrero98, Melic98, Blinov21} have also been obtained. Thus, it is of immense interest to obtain explicit kink solutions in as many higher order field theory models as possible.

The purpose of this paper is to partially construct some of the exact solutions. In particular, we present four one-parameter families of higher order field theory models for which we are able to obtain explicit kink solutions. Two of these models have a kink solution with an exponential tail while the other two admit kink solutions with a power-law tail. To the best of our knowledge, these are the first known examples of a one-parameter family of higher order field theory models admitting kink solutions with a power-law tail. We study in detail various properties of these kink solutions including the kink mass, nature of the kink tail, the kink linear stability analysis and offer predictions for the kink-kink (KK) and the kink-antikink (K-AK) forces by using the recent Manton formalism \cite{Manton19} and its extension to higher order field theories \cite{ChristovPRL}. For the one-parameter family with an exponential tail, the calculation of the KK and K-AK forces is straightforward using the celebrated Manton formalism \cite{Manton79}.

The plan of the paper is as follows. In Sec. II we obtain explicit kink 
solutions in one-parameter families of potentials of the type 
$\phi^2(a^{2n} - \phi^{2n})^2(b^{2n} - \phi^{2n})^2$ in case $b$ and $a$ are 
related to each other. We study in detail various properties of these kink 
solutions all of which have an exponential tail. In Secs. III and IV we further
extend these results. In Sec. III we obtain 
explicit kink solutions in a one-parameter family of potentials of the type $\phi^2(a^{2n} - \phi^{2n})^2(b^{2n} - \phi^{2n})^2(c^{2n} - \phi^{2n})^2$ in case $c$, $b$ and $a$ are related to each
other.  We study in detail the properties of these kink solutions all of which have an exponential kink tail.  Section IV contains yet another family of solutions with the potential $V(\phi) = \lambda^2 \phi^{2} (a^{2n}-\phi^{2n})^2 (b^{2n}+\phi^{2n})^2$. 
In the next two sections we study one-parameter families of higher order field theory models and obtain explicit kink solutions with a power-law tail. 
In particular, in Sec. V we obtain explicit kink solutions in a one-parameter family of potentials of the type
$\phi^{2n+2}|a^{2n} - \phi^{2n}|^{3/2}$ all of which admit kink solutions with 
a power law-tail. We study the properties
of these solutions including the stability analysis and show that in this case, as expected, there is no gap between the zero mode and the beginning of the continuum. By using the recent Manton formalism \cite{Manton19} we provide predictions for the K-K, AK-K and K-AK acceleration as well as the ratio of the AK-K and K-K forces. In Sec. VI we present another one-parameter family
of potentials of the form $|a^{2n}  - \phi^{2n}|^{(2n+1)/2n}$ and obtain explicit kink solutions with a power-law tail. In this case too, we study in detail the various properties of the
kink solutions and make predictions for the K-K and AK-K acceleration as well 
as the ratio of the corresponding forces. Finally, in Sec. VII we summarize 
the results obtained in this paper and point out some of the open problems.  
In Appendices A and B we provide few more explicit kink solutions which are
obtained by solving quartic equations. Further, for completeness, we also
briefly discuss known explicit kink solutions in Appendix C.

\section{Explicit Kink Solutions with Exponential Tail for a One-Parameter Family of Potentials} 

Let us consider the one-parameter family of potentials as given by 
\be\label{1}
V(\phi) = \lambda^2 \phi^2 (a^{2n}-\phi^{2n})^2 (b^{2n}-\phi^{2n})^2\,,~~
n=1,2,3,..., ~~b > a>0\,.
\ee
Note that this potential has 5 degenerate minima at $\phi = 0, \pm a, \pm b$ (see Fig.~1) 
and hence two kink and two mirror kink solutions and corresponding four
antikink solutions. We now obtain these kink solutions explicitly. 

\begin{figure}[h] 
\includegraphics[width=6.0 in]{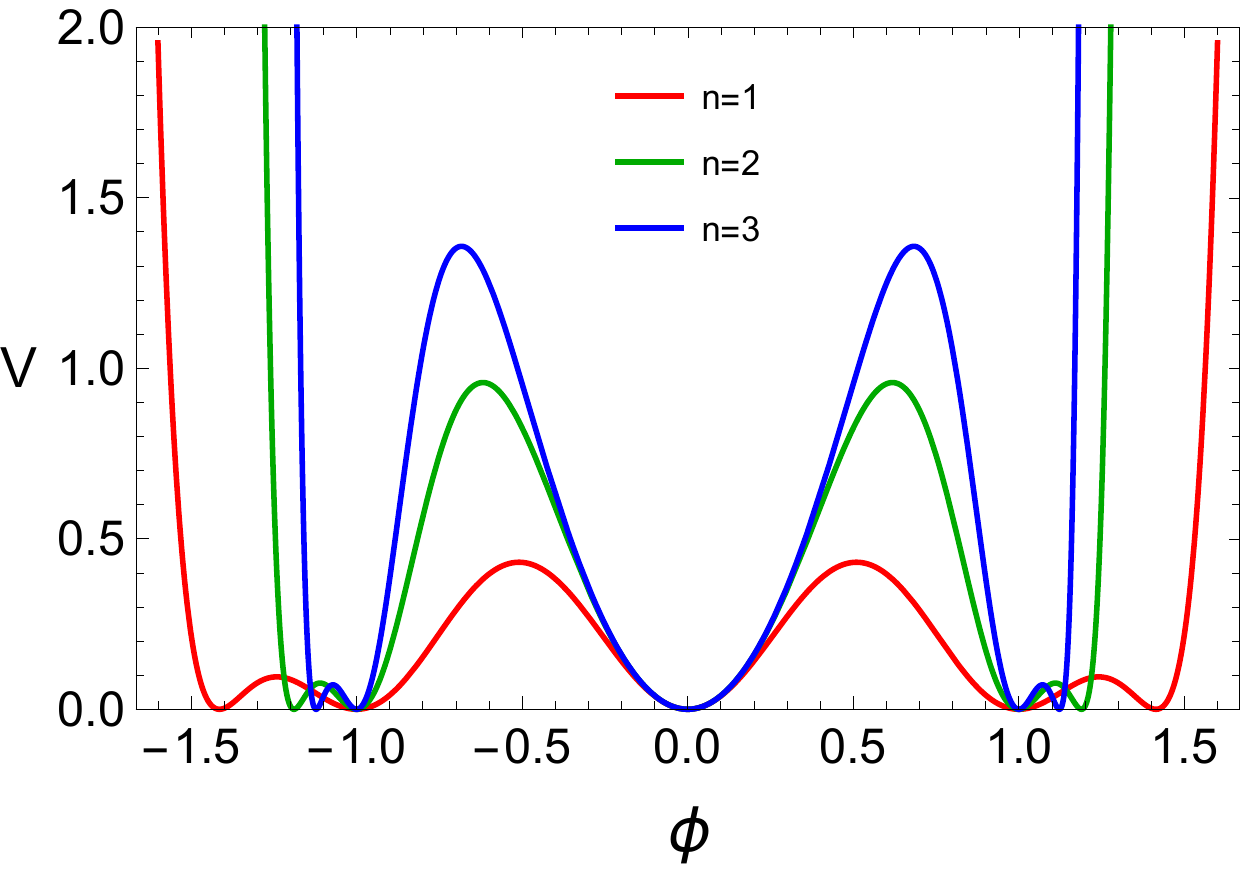}
\caption{Potential $V(\phi)$ with five degenerate minima for three different values of $n=1,2,3$.  Parameters are $\lambda=1$, $a=1$ and $b=2^{1/(2n)}$. See Eq. (1).  }
\end{figure} 

\subsection{Kink Solution From $0$ to $a$}

In this case, we need to solve the self-dual equation
\be\label{2}
\frac{d\phi}{dx} = \lambda \phi (a^{2n} -\phi^{2n}) (b^{2n} -\phi^{2n})\,, 
\ee
which can be achieved by using the partial fraction
\be\label{3}
\int d\phi\, \bigg [\frac{A}{\phi} +\frac{B\phi^{2n-1}}{(a^{2n}- \phi^{2n})}
+\frac{D \phi^{2n-1}}{(b^{2n} -\phi^{2n})} \bigg ]\,.
\ee
It is straightforward to check that 
\be\label{4}
A = \frac{1}{a^{2n} b^{2n}}\,,~~B = \frac{1}{a^{2n}(b^{2n}-a^{2n})}\,,~~
D = -\frac{1}{b^{2n}(b^{2n}-a^{2n})}\,.
\ee
Eq. (\ref{3}) is then easily integrated giving an implicit kink solution
\be\label{5}
e^{\mu x} = \frac{(\phi^{2n})^{(b^{2n}-a^{2n})} (b^{2n}-\phi^{2n})^{a^{2n}}}
{(a^{2n}-\phi^{2n})^{b^{2n}}}\,,
\ee
where
\be\label{6}
\mu = 2\sqrt{2} n \lambda a^{2n} b^{2n} (b^{2n}-a^{2n})\,.
\ee
Note that the implicit kink solution is valid for arbitrary values of
$a,b$ with the constraint $b > a > 0$.

It is straightforward to calculate the kink mass in this case. We find
\be \label{6a}
M_K = \int_{0}^{a} d\phi\, \sqrt{2V(\phi)} 
= \frac{\lambda n a^{2(n+1)}[(2n+1)b^{2n} - a^{2n}]}{\sqrt{2} (2n+1)(n+1)}\,.
\ee

We now consider three special cases in which Eq. (\ref{5}) can be easily
inverted yielding an explicit kink solution. Two of these cases we discuss
below while the third case is discussed in Appendix A.

\subsubsection{Case I: $b^{2n} = 2a^{2n}$} 

In this case the implicit kink solution takes the form
\be\label{7}
e^{\mu p x} = \frac{\phi^{2n} (2a^{2n}-\phi^{2n})}{(a^{2n}-\phi^{2n})^2}\,,
\ee
where $p$ is an arbitrary positive number with $pa^{2n} =1$, while $\mu$ now has a simple form
\be\label{8}
\mu = \frac{4\sqrt{2} n \lambda}{p^3}\,.
\ee
Eq. (\ref{7}) is easily inverted yielding an explicit kink solution from 
$0$ to $a$ (see Fig. 2) 
\be\label{9}
\phi_{K}(x) = a \left[1-\frac{1}{\sqrt{1+e^{\mu p x}}}\right]^{1/2n}\,.
\ee
Thus while the implicit kink solution is known for arbitrary values of 
$a,b$ satisfying $b > a > 0$, the explicit kink solution while valid 
for arbitrary values of $a$, the corresponding $b$ is however fixed, i.e. 
$b^{2n} = 2 a^{2n}$.   

\begin{figure}[h] 
\includegraphics[width=6.0 in]{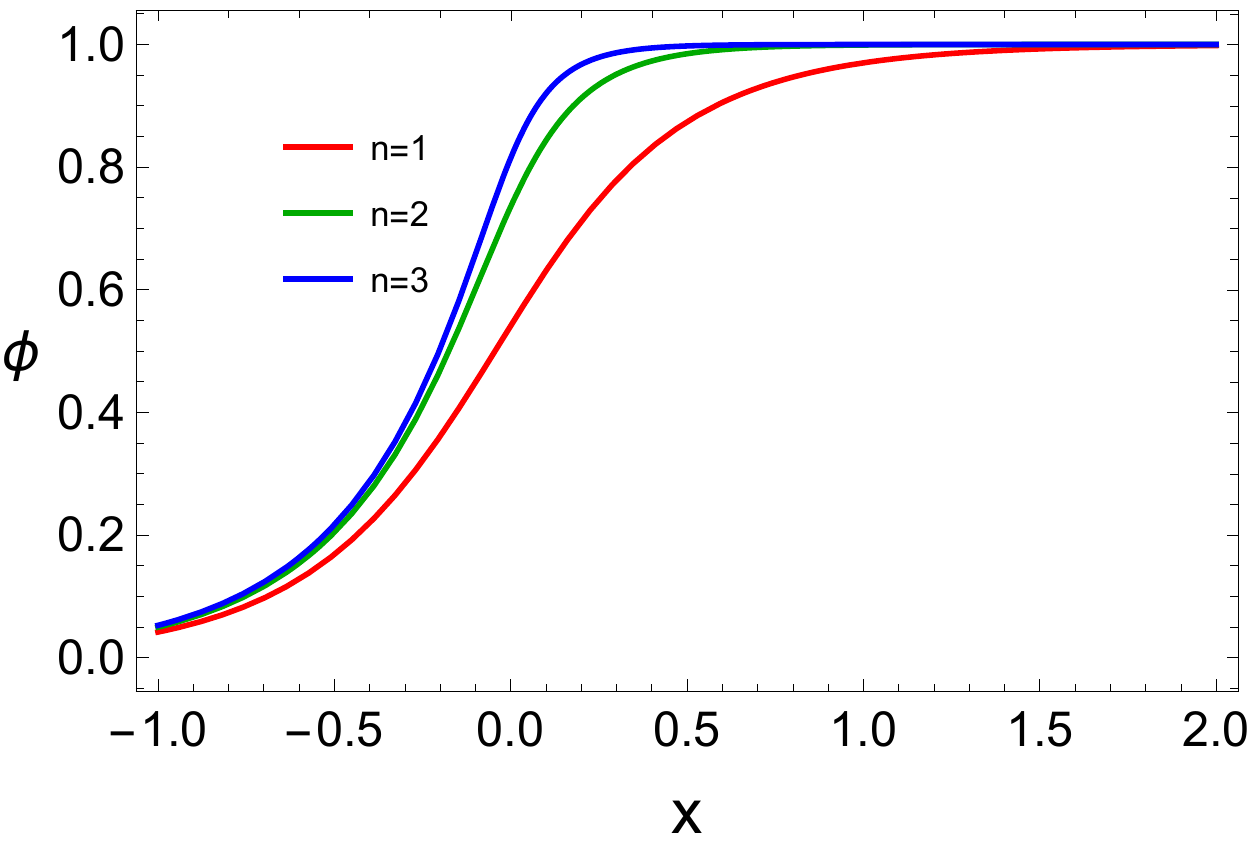}
\caption{Kink solution $\phi(x)$ given by Eq. (10) for $a=1$, $p=1$ and $\mu=4\sqrt{2}n$. }
\end{figure} 

It is straightforward to check that
\be\label{9a}
\lim_{x \rightarrow -\infty} \phi(x) = 	a 2^{-1/2n} e^{\mu p x/2n}\,,~~
\lim_{x \rightarrow \infty} \phi(x) = 	a-(2n)^{-1} a e^{-\mu p x/2}\,.
\ee
Further, the zero mode in this case is easily calculated
\be\label{9b}
\psi^{0,K} \propto \frac{e^{\mu p x}}{(1+e^{\mu p x})^{3/2}} 
\frac{1}{[1-(1+e^{\mu p x})^{-1/2}]^{1-1/2n}}\,.
\ee
Finally, it is straightforward to carry out kink stability analysis. The kink
stability potential 
$V_{K}(x)$ which appears in the 
Schr\"odinger-like equation 
\be\label{9s} 
-\frac{d^2\psi}{dx^2} + V_K(x)\psi = \omega^2 \psi \,, 
\ee
is easily calculated
\bea\label{9c}
&&V(x) = \frac{d^2V(\phi)}{d\phi^{2}}\Bigg|_{\phi = \phi_K(x)} \nonumber \\
&& = 2\lambda^2 \bigg [(a^{2n}-\phi_{K}^{2n})^2 (2a^{2n}-\phi_{K}^{2n})^2 
+16 n^2 \phi_{K}^{4n}(a^{2n}-\phi_{K}^{2n}) (2a^{2n}-\phi_{K}^{2n}) \nonumber \\
&&+4n^2 \phi_{K}^{4n}[(2a^{2n}-\phi_{K}^{2n})^2 + (a^{2n}-\phi_{K}^{2n})^2] 
\nonumber \\
&&-2n(2n+3)\phi_{K}^{2n}(a^{2n}-\phi_{K}^{2})(2a^{2n}-\phi_{K}^{2}) 
(3a^{2n}-2\phi_{K}^{2n}) \bigg ]\,,
\eea
and substituting the kink solution (\ref{9}) in it. In particular, 
since the kink solution goes from $0$ to $a$ as $x$ goes
from $-\infty$ to $\infty$, it is easy to see that 
\bea\label{9d} 
&&V(x = \infty) = \frac{8\lambda n^2}{p^4}\,,~~V(x = -\infty) 
= \frac{8\lambda}{p^4}\,, \nonumber \\
&&V(x = 0)
=\frac{[(3+2\sqrt{2})/4 -(\sqrt{2}-1) n^2 -(3+\sqrt{2})n]}{p^4}\,.
\eea
It will be interesting to study if this potential admits other discrete modes
than the zero mode.

\subsubsection{Case II: $b^{2n} = 3a^{2n}$} 

In this case the implicit kink solution takes the form
\be\label{10}
e^{\mu q x} = \frac{\phi^{4n} (3a^{2n}-\phi^{2n})}{(a^{2n}-\phi^{2n})^3}\,,
\ee
where $q$ is any positive number satisfying $a^{2n} q = 1$ and $\mu$ now has 
a simple form
\be\label{11}
\mu = \frac{12\sqrt{2} n \lambda}{q^3}\,.
\ee
Eq. (\ref{10}) is easily inverted yielding a cubic equation
\be\label{12}
u^3 +\frac{3u}{(e^{\mu q x}-1)}-\frac{2}{(e^{q \mu x}-1)} = 0\,,
\ee
where $u = 1 - (\frac{\phi}{a})^{2n}$. Thus the explicit kink solution is 
given by
\bea\label{13}
u = 1-\left(\frac{\phi}{a}\right)^{2n} = \bigg [\frac{(\sqrt{e^{\mu q x}-1)}
+e^{\mu q x /2}}
{(e^{\mu q x}-1)^{3/2}} \bigg]^{1/3}  
+\bigg [\frac{\sqrt{(e^{\mu q x}-1)}-e^{\mu q x /2}}
{(e^{\mu q x}-1)^{3/2}} \bigg ]^{1/3}\,.
\eea   
It is straightforward to check that
\be\label{13a}
\lim_{x \rightarrow -\infty} \phi(x) = 	3^{-1/4n} a e^{\mu q x/4n}\,,~~
\lim_{x \rightarrow \infty} \phi(x) = 	a-\frac{a}{2^{2/3}n} e^{-\mu q x/3}\,.
\ee
Thus while the implicit kink solution is known for arbitrary values of 
$a,b$ satisfying $b > a > 0$, the explicit kink solution (\ref{13}) while valid 
for arbitrary values of $a$, the corresponding $b$ is however fixed, i.e. 
$b^{2n} = 3 a^{2n}$.   

Proceeding in the same way, by choosing $b^{2n} = m a^{2n}$ one can in 
principle obtain an explicit kink solution by solving an $m$'th order 
equation.

\subsection{Kink Solution From $a$ to $b$}

In this case, we need to solve the self-dual equation
\be\label{14}
\frac{d\phi}{dx} = \lambda \phi (a^{2n} -\phi^{2n}) (b^{2n} -\phi^{2n})\,, 
\ee
which is easily achieved by using the partial fraction
\be\label{15}
\int d\phi\, \bigg [\frac{A}{\phi} +\frac{B\phi^{2n-1}}{(\phi^{2n} a^{2n})}
+\frac{D \phi^{2n-1}}{(b^{2n} -\phi^{2n})} \bigg ]\,.
\ee
It is straightforward to check that 
\be\label{16}
A = -\frac{1}{a^{2n} b^{2n}}\,,~~B = \frac{1}{a^{2n}(b^{2n}-a^{2n})}\,,~~
D = -\frac{1}{b^{2n}(b^{2n}-a^{2n})}\,.
\ee
Eq. (\ref{15}) is then easily integrated giving an implicit kink solution
\be\label{17}
e^{-\mu x} = \frac{(\phi^{2n})^{(b^{2n}-a^{2n})} (b^{2n}-\phi^{2n})^{a^{2n}}}
{(a^{2n}-\phi^{2n})^{b^{2n}}}\,,
\ee
where
\be\label{18}
\mu = 2\sqrt{2} n \lambda a^{2n} b^{2n} (b^{2n}-a^{2n})\,.
\ee
It is straightforward to calculate the kink mass in this case. We find
\bea\label{18a}
M_K = \int_{a}^{b} d\phi\, \sqrt{2V(\phi)} 
= \frac{\lambda n a^{2}}{\sqrt{2}} \bigg [\frac{([b/a]^{4n+2}-1)}{(n+1)(n+2)}
-\frac{(b/a)^{2n}([b/a]^2 -1)}{(n+1)} \bigg ]\,.
\eea

We now consider three special cases in which Eq. (\ref{17}) can be easily
inverted yielding an explicit kink solution, two given below and the third in Appendix A.

\subsubsection{Case I: $b^{2n} = 2a^{2n}$} 

In this case the implicit kink solution takes the form
\be\label{19}
e^{-\mu p x} = \frac{\phi^{2n} (2a^{2n}-\phi^{2n})}{(\phi^{2n}-a^{n})^2}\,,
\ee
where $p$ is an arbitrary positive number satisfying $a^{2n} p = 1$ while 
$\mu$ now has a simple form given by Eq. (9). 
Eq. (\ref{19}) is easily inverted yielding an explicit kink solution (see Fig. 3) 
\be\label{21}
\phi(x) = a \left[1+\frac{1}{\sqrt{1+e^{-\mu p x}}}\right]^{1/2n}\,.
\ee
\begin{figure}[h] 
\includegraphics[width=6.0 in]{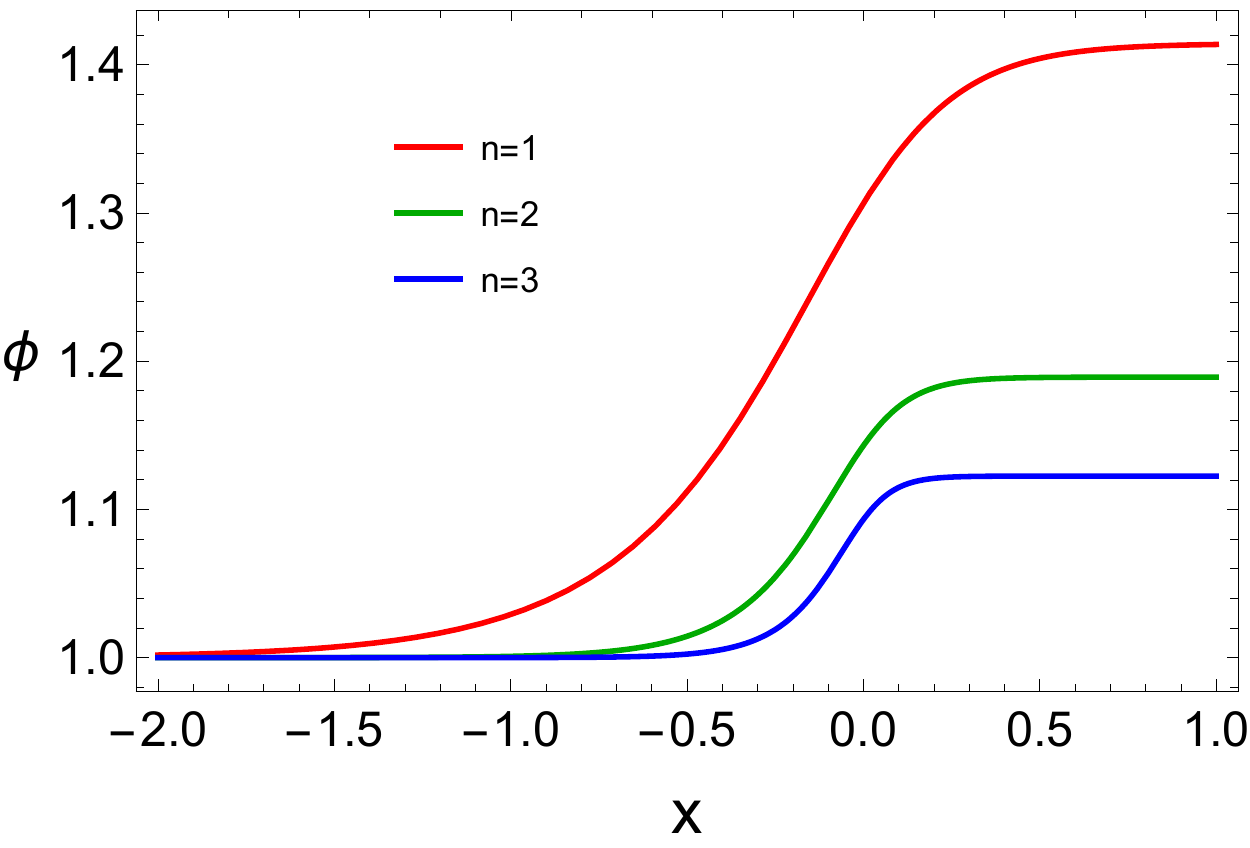}
\caption{Kink solution $\phi(x)$ given by Eq. (28) with $a=1$, $p=1$ and $\mu=4\sqrt{2}n$. }
\end{figure} 

It is straightforward to check that
\be\label{21a}
\lim_{x \rightarrow -\infty} \phi(x) = 	a+ (2n)^{-1} a e^{\mu p x/2}\,,~~
\lim_{x \rightarrow \infty} \phi(x) = 	2^{1/2n}a -(4n)^{-1} a e^{-\mu p x}\,.
\ee
Further, the zero mode in this case is easily calculated
\be\label{21b}
\psi^{0,K} \propto \frac{e^{-\mu p x}}{(1+e^{-\mu p x})^{3/2}} 
\frac{1}{[1+(1+e^{-\mu p x})^{-1/2}]^{1-1/2n}}\,.
\ee
Finally, it is straightforward to do the kink stability analysis. The kink
stability potential that appears in the Schr\"odinger like equation (\ref{9s}) 
can be easily calculated by using
\bea\label{21c}
&&V(x) = \frac{d^2V(\phi)}{d\phi^{2}}\Bigg|_{\phi = \phi_K(x)} \nonumber \\
&& = 2\lambda^2 \bigg [(\phi_{K}^{2n}-a^{2n})^2 (2a^{2n}-\phi_{K}^{2n})^2 
-16 n^2 \phi_{K}^{4n}(\phi_{K}^{2n}-a^{2n}) (2a^{2n}-\phi_{K}^{2n}) 
+4n^2 \phi_{K}^{4n}[(2a^{2n}-\phi_{K}^{2n})^2 + (\phi_{K}^{2n}-a^{2n})^2]
\nonumber \\
&&-2n(2n+3)\phi_{K}^{2n}(\phi_{K}^{2}-1)(2-\phi_{K}^{2}) 
(2\phi_{K}^{2n}-3a^{2n}) \bigg ]\,,
\eea
and substituting the kink solution (\ref{21}) in it. In particular, 
since the kink solution goes from $a$ to $2^{1/2n}a$ as $x$ goes
from $-\infty$ to $\infty$, it is easy to observe that 
\bea\label{21d} 
&&V(x = \infty) = {32\lambda n^2}{p^4}\,,~~V(x = -\infty) = 
\frac{8\lambda n^2}{p^4}\,, 
\nonumber \\
&&V(x = 0)
=\frac{[(3-2\sqrt{2})/4 -(3\sqrt{2}+2) n^2 -(3+\sqrt{2})n/2]}{p^4}\,.
\eea
For large $n$, thus $V(x = \pm \infty)$ is large and it is certain that this
stability potential has several discrete modes apart from the zero mode. In
fact in the $n \rightarrow \infty$ limit, this potential has only discrete
modes. This is rather unusual.

\subsubsection{Case II: $b^{2n} = 3a^{2n}$} 

In this case the implicit kink solution takes the form
\be\label{21e}
e^{\mu q x} = \frac{\phi^{4n} (3-\phi^{2n})}{(1-\phi^{2n})^3}\,,
\ee
where $q$ is any positive number satisfying $q a^{2n} = 1$ while $\mu$ now has 
a simple form as given by Eq. (16). 
Eq. (\ref{21e}) is easily inverted yielding a cubic equation
\be\label{24}
u^3 -\frac{3u}{(e^{-\mu q x}+1)}-\frac{2}{(e^{-\mu q x}+1)} = 0\,,
\ee
where $u = (\frac{\phi}{a})^{2n} -1$. Thus the explicit kink solution is 
given by
\bea\label{25}
u = \left(\frac{\phi}{a}\right)^{2n}-1 = \bigg [\frac{(\sqrt{e^{-\mu q x}+1)}
+e^{-\mu q x /2}}
{(e^{-\mu q x}+1)^{3/2}} \bigg ]^{1/3}  
+\bigg [\frac{\sqrt{(e^{-\mu q x}+1)}-e^{-\mu p x /2}}
{(e^{-\mu q x}+1)^{3/2}} \bigg ]^{1/3}\,.
\eea   
It is straightforward to check that
\be\label{25a}
\lim_{x \rightarrow -\infty} \phi(x) = a+a\frac{\sqrt{3}}{2n} e^{\mu q x/2}\,,~~
\lim_{x \rightarrow \infty} \phi(x) = 3^{1/2n}a -\frac{8a}{9} e^{-\mu q x}\,.
\ee

Another exact solution in this family of potentials is provided in Appendix A. 

\section{Explicit Kink Solutions with Exponential Tail for Another One-Parameter Family of Potentials} 

We now obtain exact kink solutions in yet another  one-parameter family
of higher order field theories characterized by the potential (see Fig. 4) 
\be\label{2.1}
V(\phi) = \lambda^2 \phi^2 (a^{2n}-\phi^{2n})^2 (b^{2n}-\phi^{2n})^2 
(c^{2n} - \phi^{2n})^2\,,~~ c > b > a > 0\,.
\ee
Note that this potential has 7 degenerate minima at $\phi = 0, \pm a, \pm b, 
\pm c$
and hence three kink and three mirror kink solutions and corresponding six 
antikink solutions. We now discuss the three kink solutions one by one. 

\begin{figure}[h] 
\includegraphics[width=6.0 in]{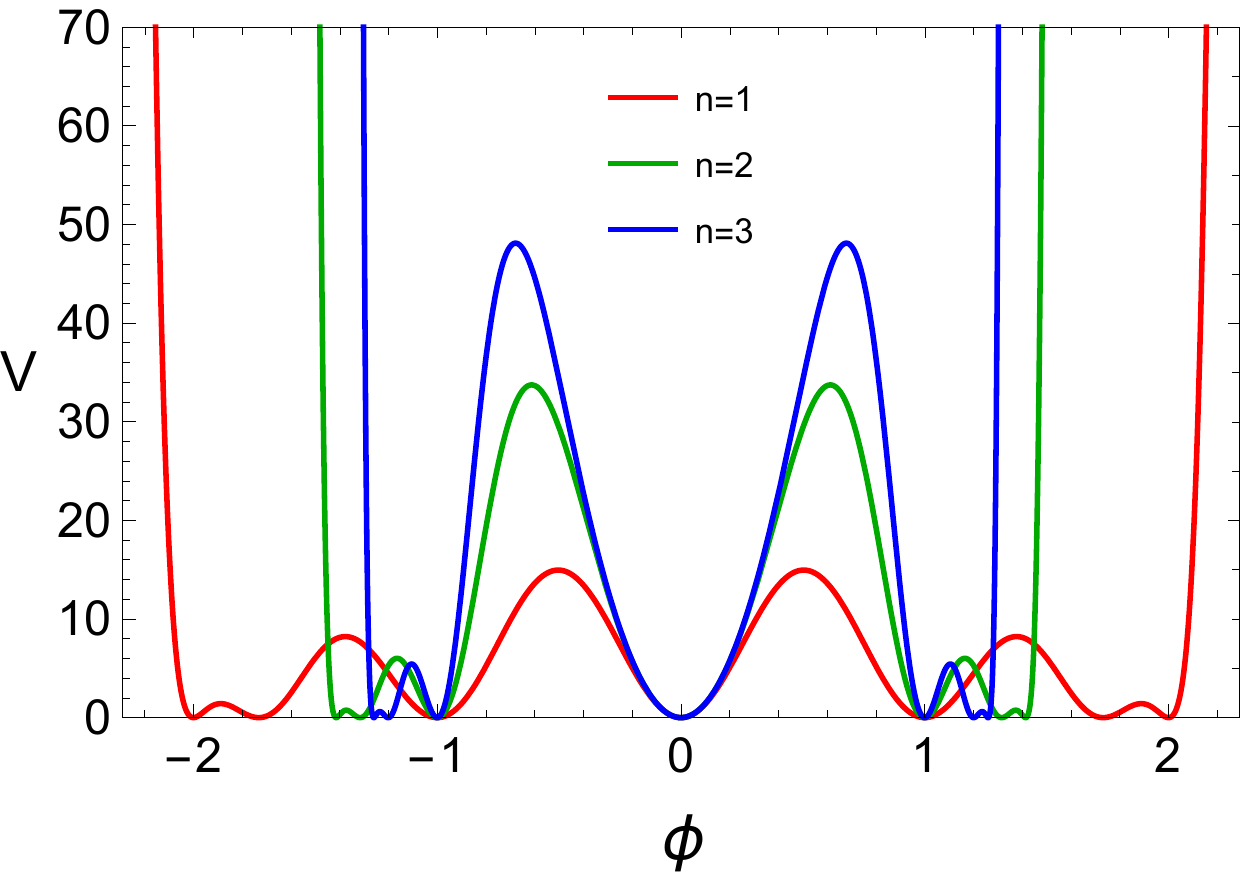}
\caption{Potential $V(\psi)$ with seven degenerate minima for three different values of $n=1,2,3$.  Parameters are $\lambda=1$, $a=1$, $b=3^{1/(2n)}$ 
and $c=4^{1/(2n)}$. See Eq. (37). }
\end{figure} 

\subsection{Kink Solution From $0$ to $a$}

In this case, we need to solve the self-dual equation
\be\label{2.2}
\frac{d\phi}{dx} = \lambda \phi (a^{2n} -\phi^{2n}) (b^{2n} -\phi^{2n}) 
(c^{2n} - \phi^{2n})\,, 
\ee
which is easily achieved by using the partial fraction
\be\label{2.3}
\int d\phi\, \bigg [\frac{A}{\phi} +\frac{B\phi^{2n-1}}{(a^{2n}- \phi^{2n})}
+\frac{D \phi^{2n-1}}{(b^{2n} -\phi^{2n})} 
+\frac{E \phi^{2n-1}}{(c^{2n} -\phi^{2n})} \bigg ]\,.
\ee
We find that 
\bea\label{2.4}
&&A = \frac{1}{a^{2n} b^{2n} c^{2n}}\,,~~~B = \frac{1}
{a^{2n} (b^{2n}-a^{2n})(c^{2n}-a^{2n})}\,, \nonumber \\
&&D = -\frac{1}{b^{2n}(b^{2n}-a^{2n})(c^{2n}-b^{2n})}\,, ~~~
E = \frac{1}{c^{2n}(c^{2n}-a^{2n})(c^{2n}-b^{2n})}\,.
\eea
Eq. (\ref{2.3}) is then easily integrated giving an implicit kink solution
\be\label{2.5}
e^{\mu x} = \frac{(\phi^{2n})^{[(b^{2n}-a^{2n})(c^{2n}-a^{2n})(c^{2n}-b^{2n})]}
(b^{2n}-\phi^{2n})^{a^{2n}c^{2n}(c^{2n}-a^{2n})}}
{(a^{2n}-\phi^{2n})^{b^{2n}c^{2n}(c^{2n}-b^{2n})}
(c^{2n}-\phi^{2n})^{b^{2n}a^{2n}(b^{2n}-a^{2n})}}\,,
\ee
where
\be\label{2.6}
\mu = 2\sqrt{2} n \lambda a^{2n} b^{2n} c^{2n} (b^{2n}-a^{2n}) (c^{2n}-a^{2n})
(c^{2n}-b^{2n})\,.
\ee
Note that the implicit kink solution is valid for arbitrary values of
$a, b, c$ with the constraint $c > b > a > 0$.

It is straightforward to calculate the kink mass in this case. We find
\bea\label{2.7}
M_K = \int_{0}^{a} d\phi\, \sqrt{2V(\phi)} 
= \frac{\lambda n a^{2(3n+1)}[6 \alpha \beta n^2 +(5\alpha \beta +2\alpha +
2\beta+1)n +\alpha \beta - \alpha -\beta -1]}{\sqrt{2} (3n+1)(2n+1)(n+1)}\,,
\eea
where 
\be\label{2.8a}
\alpha = \frac{b^{2n}}{a^{2n}}\,,~~\beta = \frac{c^{2n}}{a^{2n}}\,.
\ee

We now consider a special case in which Eq. (\ref{2.5}) can be easily
inverted yielding an explicit kink solution.

\subsubsection{Special Case: $b^{2n} = 3a^{2n}, ~c^{2n} = 4a^{2n}$} 

In this case the implicit kink solution takes the form
\be\label{2.8}
e^{\mu p x} = \frac{\phi^{2n} (3a^{2n}-\phi^{2n})^2}
{(a^{2n}-\phi^{2n})^2 (4a^{2n}-\phi^{2n})}\,,
\ee
where $p$ is an arbitrary positive number with $6pa^{6n} =1$, while $\mu$ now has a simple form
\be\label{2.9}
\mu = \frac{2\sqrt{2} n \lambda}{p^2}\,.
\ee
Eq. (\ref{2.8}) is easily inverted yielding a cubic equation
\be\label{2.10}
u^3 -3 u +2 \tanh\left(\frac{\mu p x}{2}\right) = 0\,,
\ee
where $u = 2 - (\frac{\phi}{a})^{2n}$. Thus the explicit kink solution is given by (see Fig. 5) 
\be\label{2.11}
u = 2-\left(\frac{\phi}{a}\right)^{2n} = 2 \cos\big [(1/3)\cos^{-1}(-\tanh[p\mu x /2]) 
 \big ]\,,
\ee
where we have used the fact that $\cos^{-1}(1) = 0\,,~~\cos^{-1}(-1) = \pi$.
It is straightforward to check that
\be\label{2.12}
\lim_{x \rightarrow -\infty} \phi(x) = (2/3)^{1/n} a e^{\mu p x/2n}\,,~~
\lim_{x \rightarrow \infty} \phi(x) = a -	\frac{a}{\sqrt{3}n} e^{-\mu p x/2}\,.
\ee
Thus while the implicit kink solution is known for arbitrary values of 
$a,b, c$ satisfying $c > b > a > 0$, the explicit kink solution (\ref{2.11}),
 while valid 
for arbitrary values of $a$, the corresponding $b, c$ are however fixed, i.e. 
$b^{2n} = 3 a^{2n}, c^{2n} = 4a^{2n}$.   

\begin{figure}[h] 
\includegraphics[width=6.0 in]{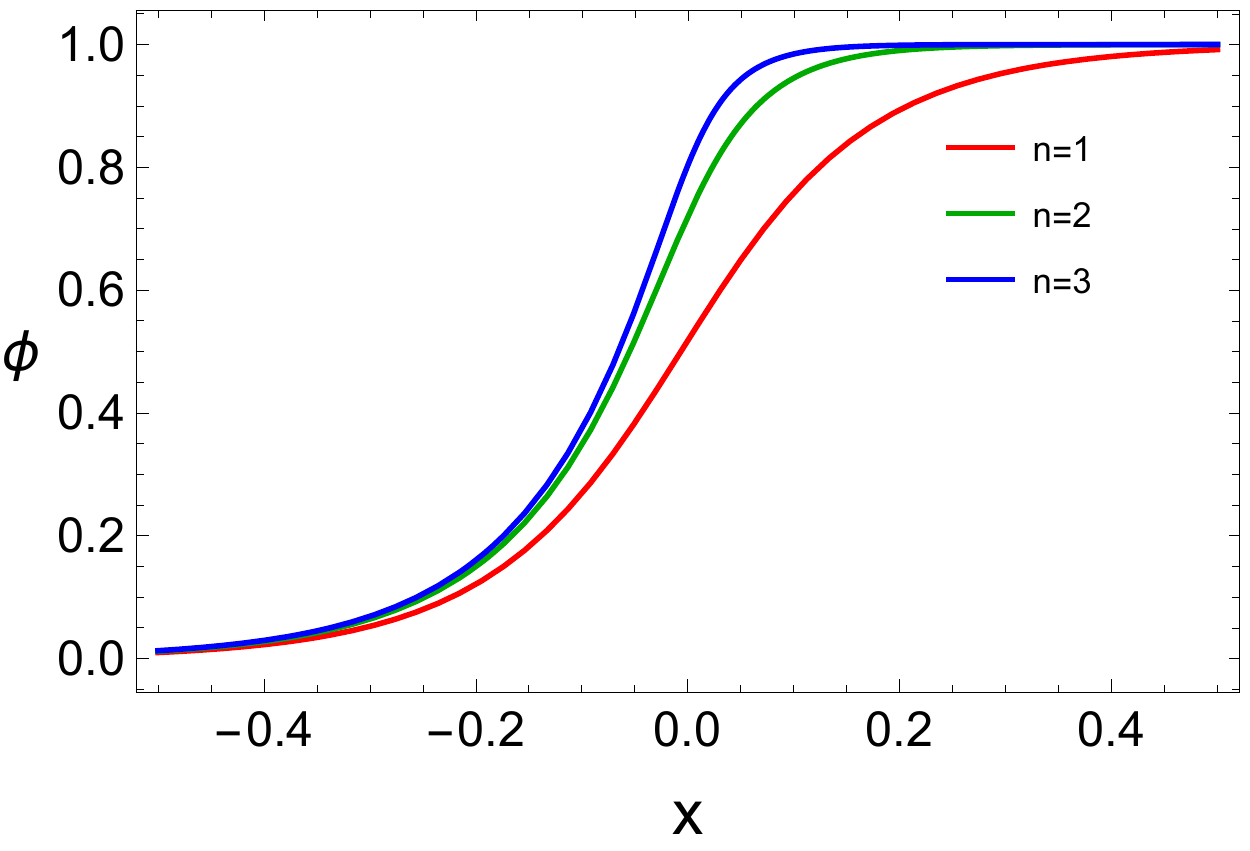}
\caption{Kink solution $\phi(x)$ given by Eq. (48) for $a=1$, $p=1/6$ and $\mu=72\sqrt{2}n$. }
\end{figure} 

\subsection{Kink Solution From $a$ to $b$}

In this case, we need to solve the self-dual equation
\be\label{2.13}
\frac{d\phi}{dx} = \lambda \phi (\phi^{2n}-a^{2n}) (b^{2n} -\phi^{2n}) 
(c^{2n} - \phi^{2n})\,, 
\ee
which is easily achieved by using the partial fraction
\be\label{2.14}
\int d\phi\, \bigg [\frac{A}{\phi} +\frac{B\phi^{2n-1}}{(\phi^{2n}- a^{2n})}
+\frac{D \phi^{2n-1}}{(b^{2n} -\phi^{2n})} 
+\frac{E \phi^{2n-1}}{(c^{2n} -\phi^{2n})} \bigg ]\,.
\ee
We find 
\bea\label{2.15}
&&A = -\frac{1}{a^{2n} b^{2n} c^{2n}}\,,~~~B = \frac{1}
{a^{2n}(b^{2n}-a^{2n})(c^{2n}-a^{2n})}\,, \nonumber \\
&&D = \frac{1}{b^{2n}(b^{2n}-a^{2n})(c^{2n}-b^{2n})}\,, ~~~ 
E = -\frac{1}{c^{2n}(c^{2n}-a^{2n})(c^{2n}-b^{2n})}\,.
\eea
Eq. (\ref{2.14}) is then easily integrated giving an implicit kink solution
\be\label{2.16}
e^{\mu x} = \frac{(\phi^{2n}-a^{2n})^{b^{2n}c^{2n}(c^{2n}-b^{2n})}
(c^{2n}-\phi^{2n})^{b^{2n}a^{2n}(b^{2n}-a^{2n})}}
{(\phi^{2n})^{[(b^{2n}-a^{2n})(c^{2n}-a^{2n})(c^{2n}-b^{2n})]}
(b^{2n}-\phi^{2n})^{a^{2n}c^{2n}(c^{2n}-a^{2n})}}\,,
\ee
where $\mu$ is given by Eq. (\ref{2.6}). 
Note that the implicit kink solution is valid for arbitrary values of
$a, b, c$ with the constraint $c > b > a > 0$.

It is straightforward to calculate the kink mass in this case. We find
\bea\label{2.18}
 M_K = \int_{a}^{b} d\phi\, \sqrt{2V(\phi)} 
&=& \frac{\lambda n a^{2(3n+1)}}{\sqrt{2}(3n+1)(2n+1)(n+1)}
\bigg [ \big (6 \alpha \beta n^2 +[5\alpha \beta +2\alpha +
2\beta+1]n +\alpha \beta - \alpha -\beta -1 \big ) \nonumber \\
&+& \alpha^{(n+1)/n} \big (-6\beta n^2 +[3\alpha \beta + 3\alpha-5\beta
-\alpha^2]n +\alpha \beta +\alpha -\alpha^2 -\beta \big ) \bigg ]\,, 
\eea
where $\alpha, \beta$ are as given by Eq. (\ref{2.8a}). 

We now consider a special case in which Eq. (\ref{2.16}) can be easily
inverted yielding an explicit kink solution.

\subsubsection{Special Case: $b^{2n} = 3a^{2n}, ~c^{2n} = 4a^{2n}$} 

In this case the implicit kink solution takes the form
\be\label{2.20}
e^{\mu p x} = \frac{(\phi^{2n}-a^{2n})^2 (4a^{2n}-\phi^{2n})}
{\phi^{2n} (3a^{2n}-\phi^{2n})^2} \,, 
\ee
where p is any positive number with $6pa^{6n} =1$, while $\mu$ now has a simple form as given by 
Eq. (\ref{2.9}).
Eq. (\ref{2.20}) is easily inverted yielding a cubic equation
\be\label{2.22}
u^3 -3 u +2 \tanh\left(\frac{\mu p x}{2}\right) = 0\,,
\ee
where $u = (\frac{\phi}{a})^{2n} -2$. Thus the explicit kink solution is given by (see Fig. 6) 
\be\label{2.23}
u = \left(\frac{\phi}{a}\right)^{2n} -2 
= 2 \cos \big [(1/3)\cos^{-1}(-\tanh[p\mu x /2]) -2\pi/3\big ]\,, 
\ee
where we have used the fact that $\cos^{-1}(1) = 0\,,~~\cos^{-1}(-1) = \pi$.
It is straightforward to check that
\be\label{2.24}
\lim_{x \rightarrow -\infty} \phi(x) = a + \frac{a}{\sqrt{3}n} e^{\mu p x/2}\,,
~~\lim_{x \rightarrow \infty} \phi(x) 
= (3)^{1/2n}a - \frac{3^{1/n} a}{3\sqrt{3}n} e^{-\mu p x/2}\,.
\ee
\begin{figure}[h] 
\includegraphics[width=6.0 in]{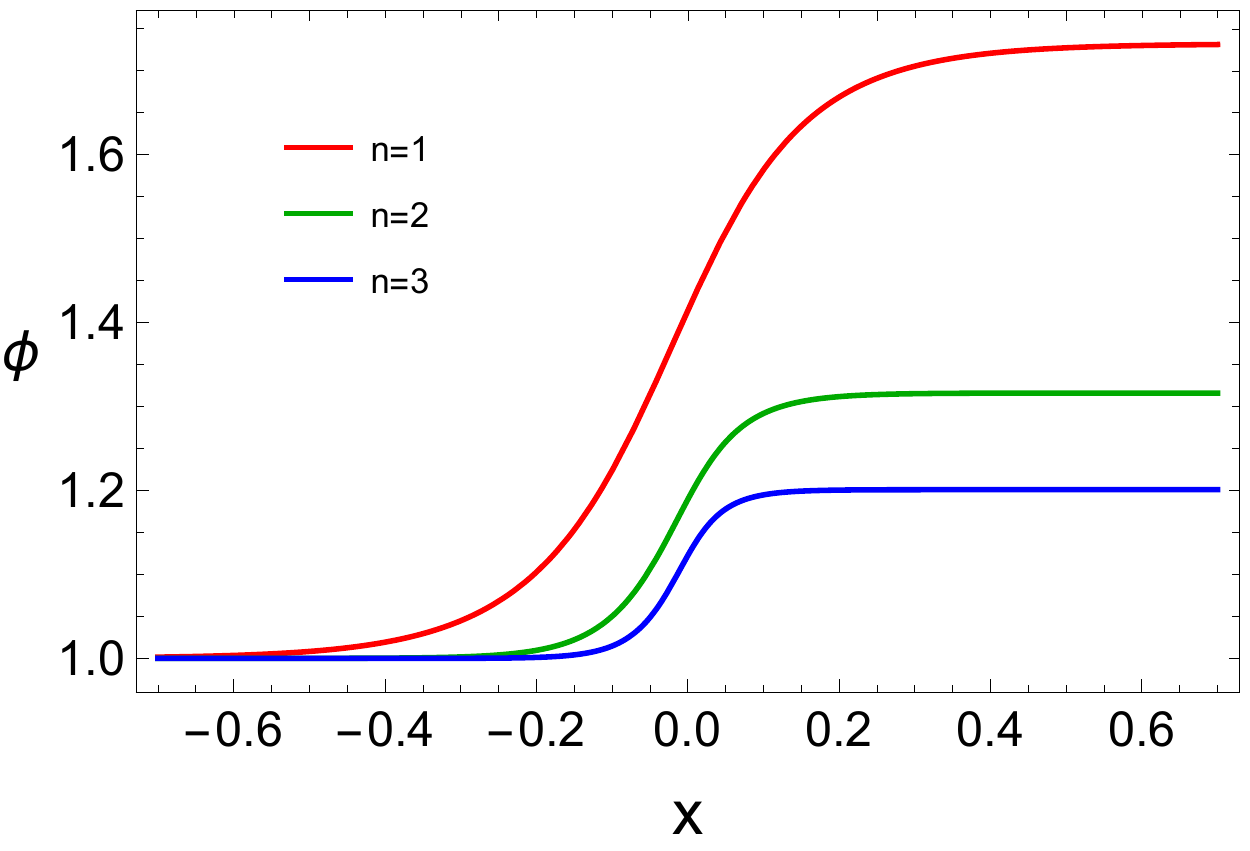}
\caption{Kink solution $\phi(x)$ given by Eq. (57) for $a=1$, $p=1/6$ and $\mu=72\sqrt{2}n$. }
\end{figure} 

\subsection{Kink Solution From $b$ to $c$}

In this case, we need to solve the self-dual equation
\be\label{2.25}
\frac{d\phi}{dx} = \lambda \phi (\phi^{2n}-a^{2n}) (\phi^{2n}-b^{2n}) 
(c^{2n} - \phi^{2n})\,, 
\ee
which is easily achieved by using the partial fraction
\be\label{2.26}
\int d\phi\, \bigg [\frac{A}{\phi} +\frac{B\phi^{2n-1}}{(\phi^{2n}- a^{2n})}
+\frac{D \phi^{2n-1}}{(\phi^{2n}-b^{2n})} 
+\frac{E \phi^{2n-1}}{(c^{2n} -\phi^{2n})} \bigg ]\,.
\ee
We find 
\bea\label{2.27}
&&A = \frac{1}{a^{2n} b^{2n} c^{2n}}\,,~~~B = -\frac{1}
{a^{2n}(b^{2n}-a^{2n})(c^{2n}-a^{2n})}\,, \nonumber \\
&&D = \frac{1}{b^{2n}(b^{2n}-a^{2n})(c^{2n}-b^{2n})}\,, ~~~ 
E = \frac{1}{c^{2n}(c^{2n}-a^{2n})(c^{2n}-b^{2n})}\,.
\eea
Eq. (\ref{2.28}) is then easily integrated giving an implicit kink solution
\be\label{2.28}
e^{\mu x} = \frac{(\phi^{2n})^{[(b^{2n}-a^{2n})(c^{2n}-a^{2n})(c^{2n}-b^{2n})]}
(\phi^{2n}-b^{2n})^{a^{2n}c^{2n}(c^{2n}-a^{2n})}}
{(\phi^{2n}-a^{2n})^{b^{2n}c^{2n}(c^{2n}-b^{2n})}
(c^{2n}-\phi^{2n})^{b^{2n}a^{2n}(b^{2n}-a^{2n})}} \,, 
\ee
where
\be\label{2.29}
\mu = 2\sqrt{2} n \lambda a^{2n} b^{2n} c^{2n} (b^{2n}-a^{2n}) (c^{2n}-a^{2n})
(c^{2n}-b^{2n})\,.
\ee
Note that the implicit kink solution is valid for arbitrary values of
$a, b, c$ with the constraint $c > b > a > 0$.

It is straightforward to calculate the kink mass in this case. We find
\bea\label{2.30}
M_K = \int_{b}^{c} d\phi\, \sqrt{2V(\phi)} 
&=& \frac{\lambda n a^{2(3n+1)}}{\sqrt{2}(3n+1)(2n+1)(n+1)} \nonumber \\
&+& \left[\beta^{(n+1)/n} \left(6\beta n^2 +[-3\alpha \beta + 5\alpha-3\beta
-\beta^2]n -\alpha \beta +\alpha +\beta^2 -\beta \right) \right]\,, 
\eea
where $\alpha, \beta$ are as given by Eq. (\ref{2.8a}). 

We now consider one special case in which Eq. (\ref{2.28}) can be easily
inverted yielding an explicit kink solution.

\subsubsection{Special Case: $b^{2n} = 3a^{2n}, ~c^{2n} = 4a^{2n}$} 

In this case the implicit kink solution takes the form
\be\label{2.32}
e^{\mu p x} = \frac{\phi^{2n} (3a^{2n}-\phi^{2n})^2}
{(\phi^{2n}-a^{2n})^2 (4a^{2n}-\phi^{2n})} \,, 
\ee
where p is any positive number with $6pa^{6n} =1$, while $\mu$ now has a simple form 
as given by Eq. (45). 
Eq. (\ref{2.32}) is easily inverted yielding a cubic equation
\be\label{2.34}
u^3 -3 u -2 \tanh\left(\frac{\mu p x}{2}\right) = 0\,,
\ee
where $u = (\frac{\phi}{a})^{2n} -2$. Thus the explicit kink solution is given 
by (see Fig. 7) 
\be\label{2.35}
u = \left(\frac{\phi}{a}\right)^{2n} -2 = 2 \cos \big [(1/3)\cos^{-1}(\tanh[p\mu x /2]) 
 \big ]\,, 
\ee
where we have used the fact that $\cos^{-1}(1) = 0\,,~~\cos^{-1}(-1) = \pi$.
It is straightforward to check that
\be\label{2.36}
\lim_{x \rightarrow -\infty} \phi(x) = (3)^{1/2n}a + \frac{a 3^{1/2n}}{3\sqrt{3}n} 
e^{\mu p x/2}\,,
~~\lim_{x \rightarrow \infty} \phi(x) 
= (2)^{1/n}a - \frac{2^{1/n} a}{18 n} e^{-\mu p x/2}\,.
\ee
\begin{figure}[h] 
\includegraphics[width=6.0 in]{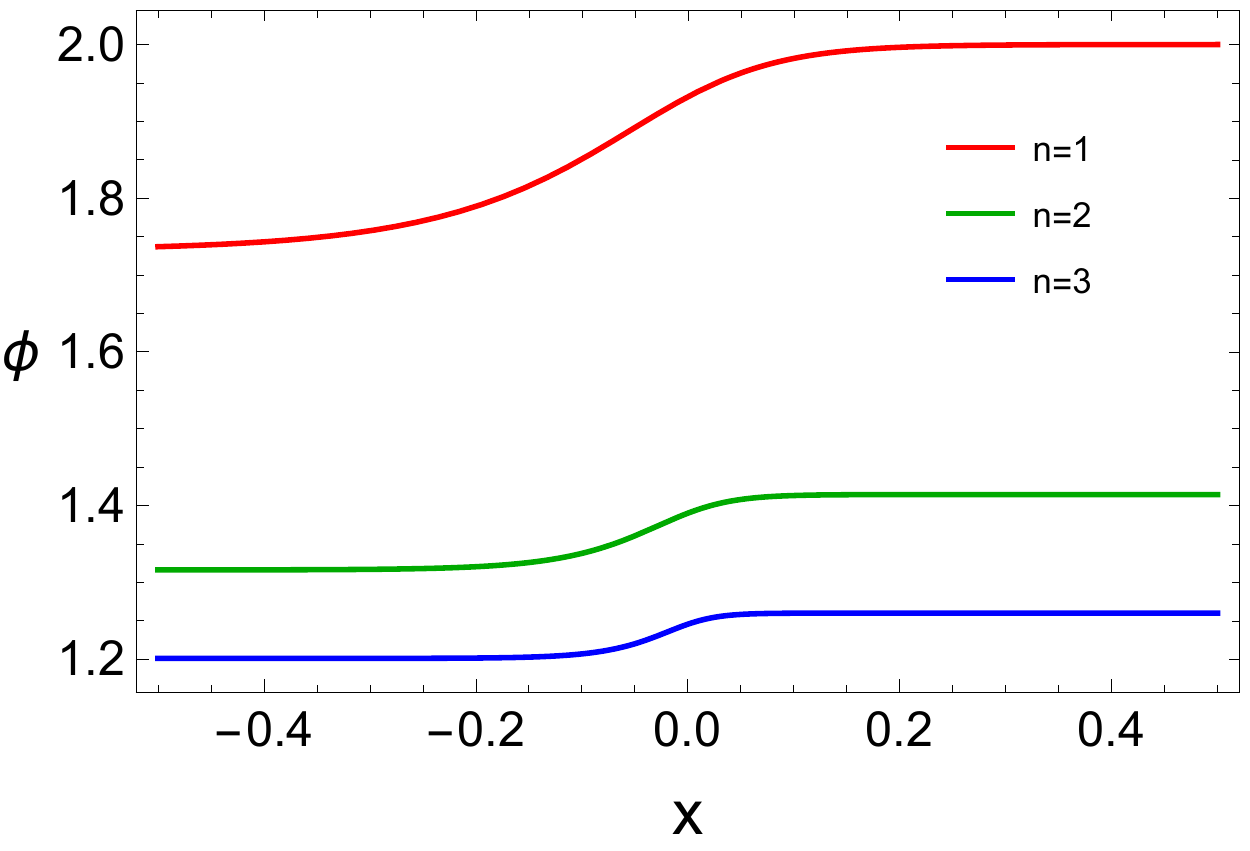}
\caption{Kink solution $\phi(x)$ given by Eq. (67) for $a=1$, $p=1/6$ and $\mu=72\sqrt{2}n$. }
\end{figure} 

\section{Analytical Kink Solutions in Another Higher Order Field Theory Model} 

We now show that one can obtain exact kink solutions in a one-parameter family
of higher order field theories characterized by the potential (see Fig. 8) 
\be\label{6.1}
V(\phi) = \lambda^2 \phi^{2} (a^{2n}-\phi^{2n})^2 (b^{2n}+\phi^{2n})^2\,.
\ee
Note that this potential has 3 degenerate minima at $\phi = 0, \pm a$
and hence one kink and one mirror kink solution and corresponding two 
antikink solutions. 

\begin{figure}[h] 
\includegraphics[width=6.0 in]{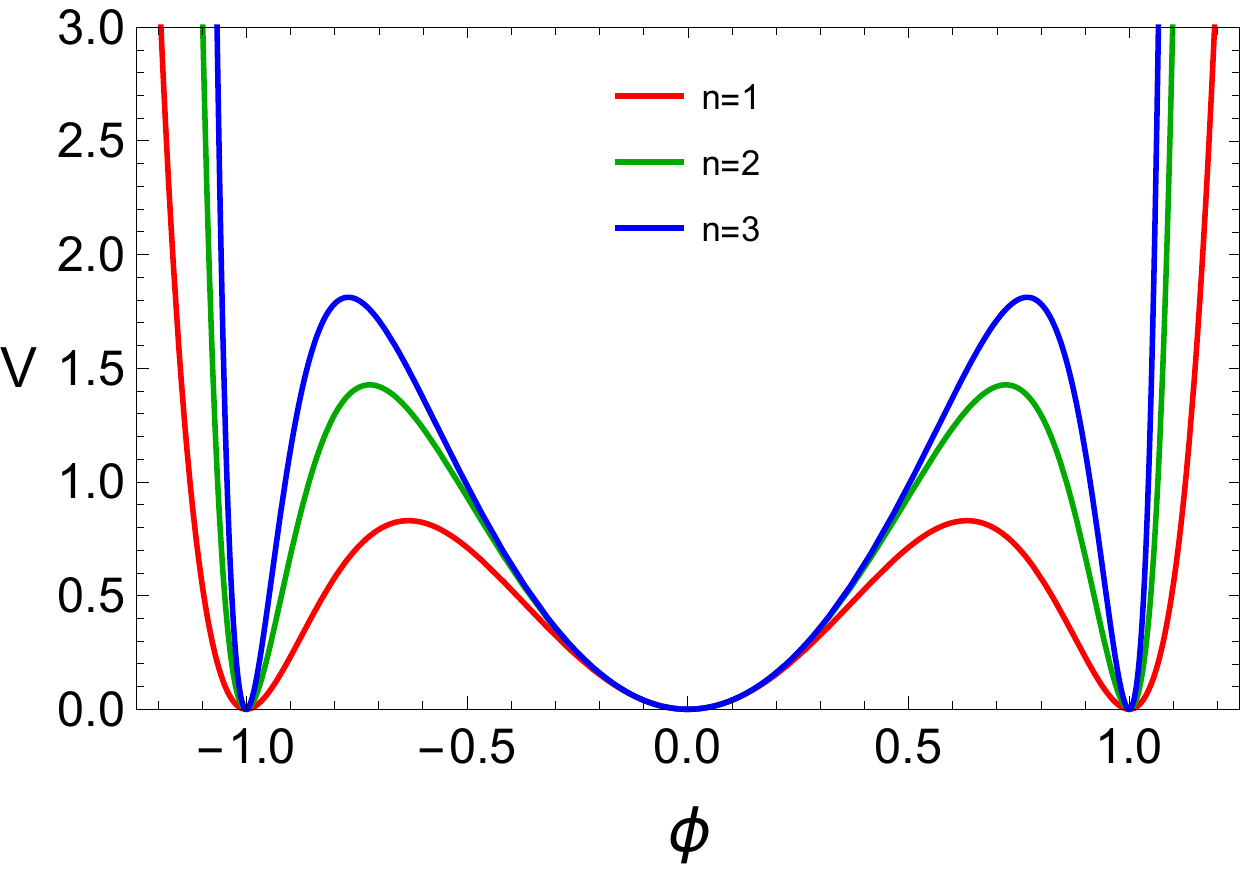}
\caption{Potential $V(\psi)$ with three degenerate minima for $n=1,2,3$. Parameters are $\lambda=1$, $a=1$ and $b=2^{1/(2n)}$. See Eq. (69) }
\end{figure} 

\subsection{\bf Kink Solution From $0$ to $a$}

In this case, we need to solve the self-dual equation
\be\label{6.2}
\frac{d\phi}{dx} = \lambda \phi (a^{2n} -\phi^{2n}) (b^{2n} +\phi^{2n})\,, 
\ee
which is easily achieved by using the partial fraction
\be\label{6.3}
\int d\phi\, \bigg [\frac{A}{\phi} +\frac{B\phi^{2n-1}}{(a^{2n}- \phi^{2n})}
+\frac{D \phi^{2n-1}}{(b^{2n} -\phi^{2n})} \bigg ]\,.
\ee
It is straightforward to check that 
\be\label{6.4}
A = \frac{1}{a^{2n} b^{2n}}\,,~~B = \frac{1}{a^{2n}(b^{2n}+a^{2n})}\,,~~
D = -\frac{1}{b^{2n}(b^{2n}+a^{2n})}\,.
\ee
Eq. (\ref{6.2}) is then easily integrated giving an implicit kink solution
\be\label{6.5}
e^{\mu x} = \frac{(\phi^{2n})^{(b^{2n}+a^{2n})}}{(b^{2n}+\phi^{2n})^{a^{2n}}
(a^{2n}-\phi^{2n})^{b^{2n}}}\,,
\ee
where
\be\label{6.6}
\mu = 2\sqrt{2} n \lambda a^{2n} b^{2n} (b^{2n}+a^{2n})\,.
\ee
Note that the implicit kink solution is valid for arbitrary values of
$a,b$.

It is straightforward to calculate the kink mass in this case. We find
\be\label{6.6a}
M_K = \int_{0}^{a} d\phi\, \sqrt{2V(\phi)} 
= \frac{\lambda n a^{2(n+1)}[(2n+1)b^{2n} + a^{2n}]}{\sqrt{2} (2n+1)(n+1)}\,.
\ee

We now consider three special cases in which Eq. (\ref{6.5}) can be easily
inverted yielding explicit kink solutions.

\subsubsection{\bf Case I: $b^{2n} = 2a^{2n}$} 

In this case the implicit kink solution (\ref{5}) takes the form
\be\label{6.7}
e^{\mu p x} = \frac{\phi^{6n}}{(2a^{2n}+\phi^{2n})(a^{2n}-\phi^{2n})^2}\,,
\ee
where $pa^{2n} =1$, while $\mu$ now has a simple form
\be\label{6.8}
\mu = \frac{12\sqrt{2} n \lambda}{p^3}\,.
\ee
Eq. (\ref{6.7}) is easily inverted yielding a cubic equation
\be\label{6.9}
(e^{p\mu x}-1)u^3 -3e^{\mu p x}u+2 e^{p \mu x} = 0\,,
\ee
where $u = (\frac{\phi}{a})^{2n}$. Notice that the solution crucially depends
on whether $x$ is positive or negative or is zero. In particular, at $x = 0$, 
$u = 2/3$, i.e. $\phi_K = (2/3)^{1/2n} a$. On the other hand the solution has
different forms in case $x> 0$ or $x < 0$. In particular, for $x > 0$ the
explicit kink solution is 
given by (see Fig. 9) 
\be\label{6.10}
u = \left(\frac{\phi_K}{a}\right)^{2n} = 2\sqrt{\frac{e^{p\mu x}}{e^{p\mu x} -1}}
\cos \bigg[\frac{1}{3} \cos^{-1} \bigg(-\sqrt{\frac{e^{p\mu x}-1}{e^{p\mu x}}}\bigg) 
\bigg ] \,,
\ee
while for $x < 0$ the solution is given by
\bea\label{6.11}
u = \left(\frac{\phi_K}{a}\right)^{2n} = \bigg [\frac{1}{1-e^{\mu p x}}
+\frac{e^{\mu p x}}
{(1-e^{\mu p x})^{3/2}} \bigg]^{1/3} 
+\bigg [\frac{1}{1-e^{\mu p x}}
-\frac{e^{\mu p x}}
{(1-e^{\mu p x})^{3/2}} \bigg]^{1/3}\,.  
\eea
It is straightforward to check that
\be\label{6.12}
\lim_{x \rightarrow -\infty} \phi_{K}(x) =	2^{1/6n} a e^{\mu p x}\,,~~
\lim_{x \rightarrow \infty} \phi_{K}(x) = 	a-\frac{a}{2\sqrt{3} n} 
e^{-\mu p x/2}\,.
\ee
Thus while the implicit kink solution is known for arbitrary values of 
$a,b$ satisfying $b > a > 0$, the explicit kink solution while valid 
for arbitrary values of $a$, the corresponding $b$ is however fixed, i.e. 
$b^{2n} = 2 a^{2n}$.   

\begin{figure}[h] 
\includegraphics[width=6.0 in]{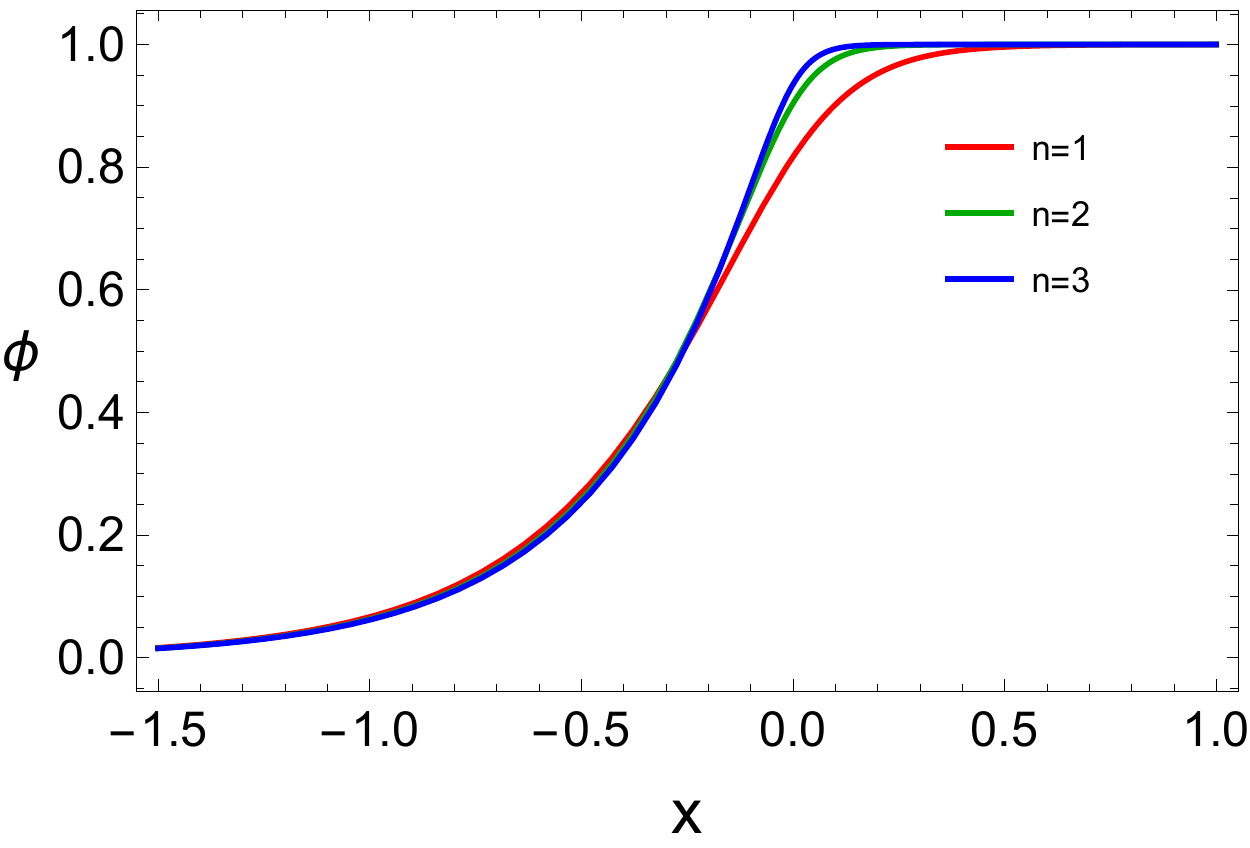}
\caption{Kink solution $\phi(x)$ given by Eqs. (79) and (80) for  $a=1$, $p=1$ and $\mu=12\sqrt{2}n$. }
\end{figure} 

It is straightforward to carry out the stability analysis by calculating the
kink potential $V_K(x) = \frac{d^2 V(\phi)}{d\phi^2}\Big|_{\phi = \phi_K}$ 
which appears in the Schr\"odinger-like equation. In particular,
we find that 
\be\label{6.12a}
V_{K}(x \rightarrow -\infty) = 8 \lambda^2 a^{8n}\,,~~~
V_{K}(x \rightarrow \infty) = 72 n^2 \lambda^2 a^{8n}\,.
\ee

\subsubsection{\bf Case II: $b^{2n} = (1/2) a^{2n}$} 

In this case the implicit kink solution (\ref{6.5}) takes the form
\be\label{6.13}
e^{2\mu p x} = \frac{\phi^{6n}}{((1/2)a^{2n}+\phi^{2n})^2 
(a^{2n}-\phi^{2n})}\,,
\ee
where $pa^{2n} =1$, while $\mu$ now has a simple form
\be\label{6.14}
\mu = \frac{3\sqrt{2} n \lambda}{2 p^3}\,.
\ee
Eq. (\ref{6.13}) is easily inverted yielding a cubic equation
\be\label{6.15}
u^3 -\frac{3e^{2\mu p x}}{4(1+e^{2p\mu x})}u
-\frac{e^{2\mu p x}}{4(1+e^{2p\mu x})} = 0\,,
\ee
where $u = (\frac{\phi}{a})^{2n}$. 
Hence the explicit kink solution is
\bea\label{6.16}
u = \left(\frac{\phi_K}{a}\right)^{2n} = \bigg [\frac{e^{2\mu p x}}
{8(1+e^{2p\mu x})}
+\frac{e^{2\mu p x}}
{16(1+e^{2\mu p x})^{3/2}} \bigg]^{1/3} 
+ \bigg [\frac{e^{2\mu p x}}
{8(1+e^{2p\mu x})}
-\frac{e^{2\mu p x}}
{16(1+e^{2\mu p x})^{3/2}} \bigg]^{1/3}\,.   
\eea
It is straightforward to check that
\be\label{6.17}
\lim_{x \rightarrow -\infty} \phi_K(x) =	2^{-1/3n} a e^{\mu p x}\,,~~
\lim_{x \rightarrow \infty} \phi_K(x) = 	a-\frac{2a}{9 n} 
e^{-2\mu p x}\,.
\ee
Thus while the implicit kink solution is known for arbitrary values of 
$a,b$, the explicit kink solution while valid 
for arbitrary values of $a$, the corresponding $b$ is however fixed, i.e. 
$b^{2n} = (1/2) a^{2n}$.   

It is straightforward to carry out the stability analysis by calculating the
kink potential $V_K(x) = \frac{d^2 V(\phi)}{d\phi^2}\Big|_{\phi = \phi_K}$ 
which appears in the Schr\"odinger-like equation (\ref{9s}). In particular,
we find that 
\be\label{6.17a}
V_{K}(x \rightarrow -\infty) =  \lambda^2 a^{8n} /2\,,~~~
V_{K}(x \rightarrow \infty) = 18 n^2 \lambda^2 a^{8n}\,.
\ee

\subsubsection{\bf Case III: $b = a$}

In this case the implicit kink solution (\ref{6.5}) is easily inverted
yielding the explicit kink solution
\be\label{6.18}
\phi = a \bigg [\frac{e^{p\mu x}}{1+e^{p \mu x}} \bigg ]^{1/4n} \,. 
\ee
It may be noted that in this case the solution is valid not only for integer $n$ but even 
half-integer $n$ and in fact has already been discussed previously by us \cite{KS19}. 
We discuss it briefly in Appendix A. 

\section{Analytical Kink Solutions with Power Law Tail in a One-Parameter 
Family of Higher Order Field Theories}

We now show that one can obtain exact kink solutions with power law tail
in a one-parameter family of higher order field theories characterized by
the potential (see Fig. 10) 
\be\label{3.1}
V(\phi) = \lambda^2 \phi^{2(n+1)} \left|(a^{2n} - \phi^{2n})\right|^{3}\,.
\ee
Note that this potential has degenerate minima at $ \phi = 0, \pm a$ and 
hence one
kink solution from $0$ to $a$ and the corresponding mirror kink solution
from $-a$ to $0$ and the corresponding two antikinks. Note that for both 
the kink and antikink solutions $-a \le \phi \le a$.  It is worth pointing out 
that whereas the potential (\ref{3.1}) is continuous, its derivative is discontinuous 
at $\phi = \pm a$.  However, since for the kink as well for the antikink 
solutions $-a \le x \le a$, this discontinuity would not matter as far as the 
kink and the antikink solutions are concerned.

\begin{figure}[h] 
\includegraphics[width=6.0 in]{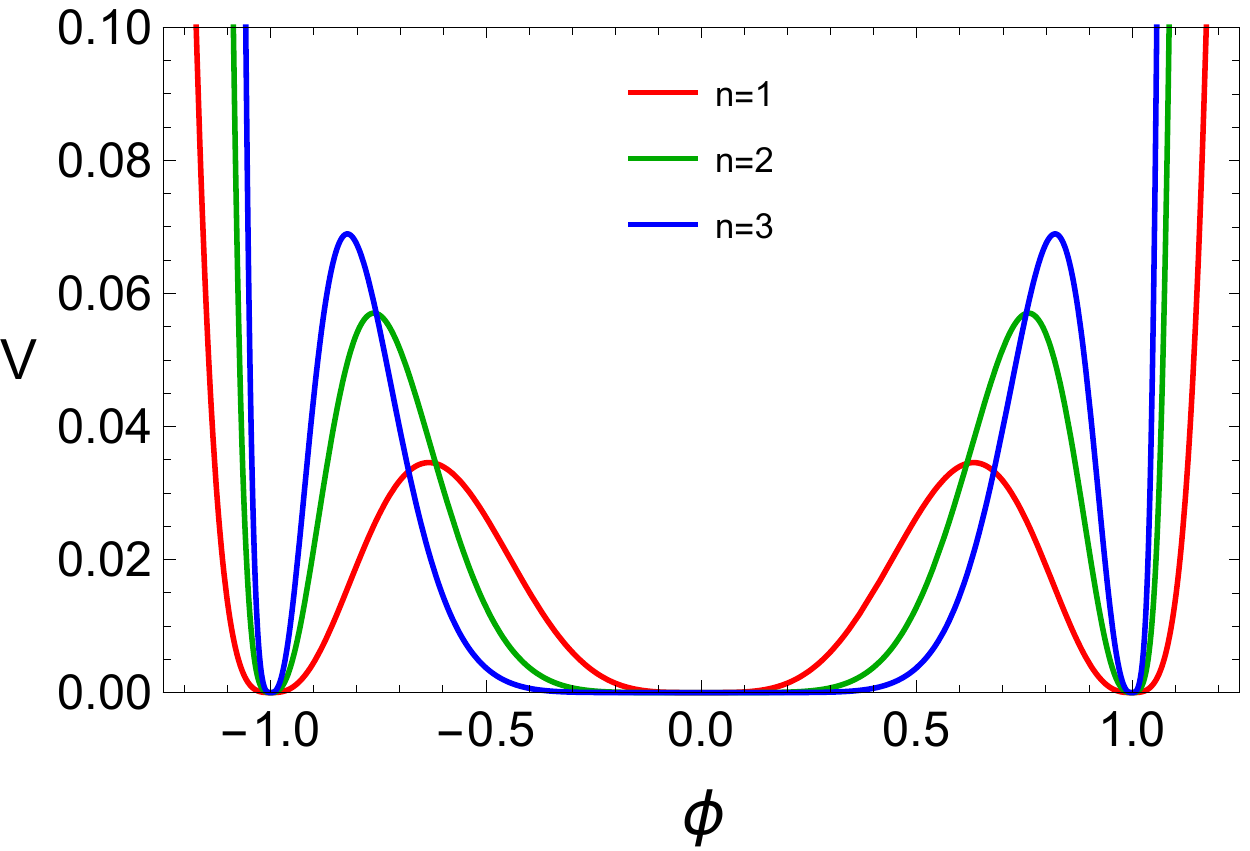}
\caption{Potential $V(\psi)$ with three degenerate minima for $n=1,2,3$.  Parameters are $\lambda=1$ and $a=1$.  See Eq. (90). }
\end{figure} 

In order to obtain the kink solution from 0 to $a$, we need to solve the self-dual
equation
\be\label{3.2}
\frac{d\phi}{dx} = \sqrt{2} \lambda \phi^{n+1} \left|\phi^{2n}-a^{2n}\right|^{3/2}\,.
\ee
This is easily integrated yielding
\be\label{3.3}
\frac{(2\phi^{2n} -a^{2n})}{\phi^{n} (a^{2n}-\phi^{2n})^{1/2}} = \mu x\,,
~~~\mu = \sqrt{2}n \lambda a^{4n}\,.
\ee
Eq. (\ref{3.3}) is easily inverted yielding an explicit kink solution (see Fig. 11) 
\be\label{3.4}
\phi_{K}(x) = \frac{a}{2^{1/2n}} \left[1+\frac{\mu x}{\sqrt{\mu^2 x^2+4}}\right]^{1/2n}\,.
\ee
It is straightforward to observe that
\be\label{3.5}
\lim_{x \rightarrow -\infty} \phi_{K}(x) =  \frac{a}{(-\mu x)^{1/n}}\,,
~~\lim_{x \rightarrow \infty} \phi_{K}(x) = a - \frac{a}{2n(\mu x)^{2}}\,.
\ee

\begin{figure}[h] 
\includegraphics[width=6.0 in]{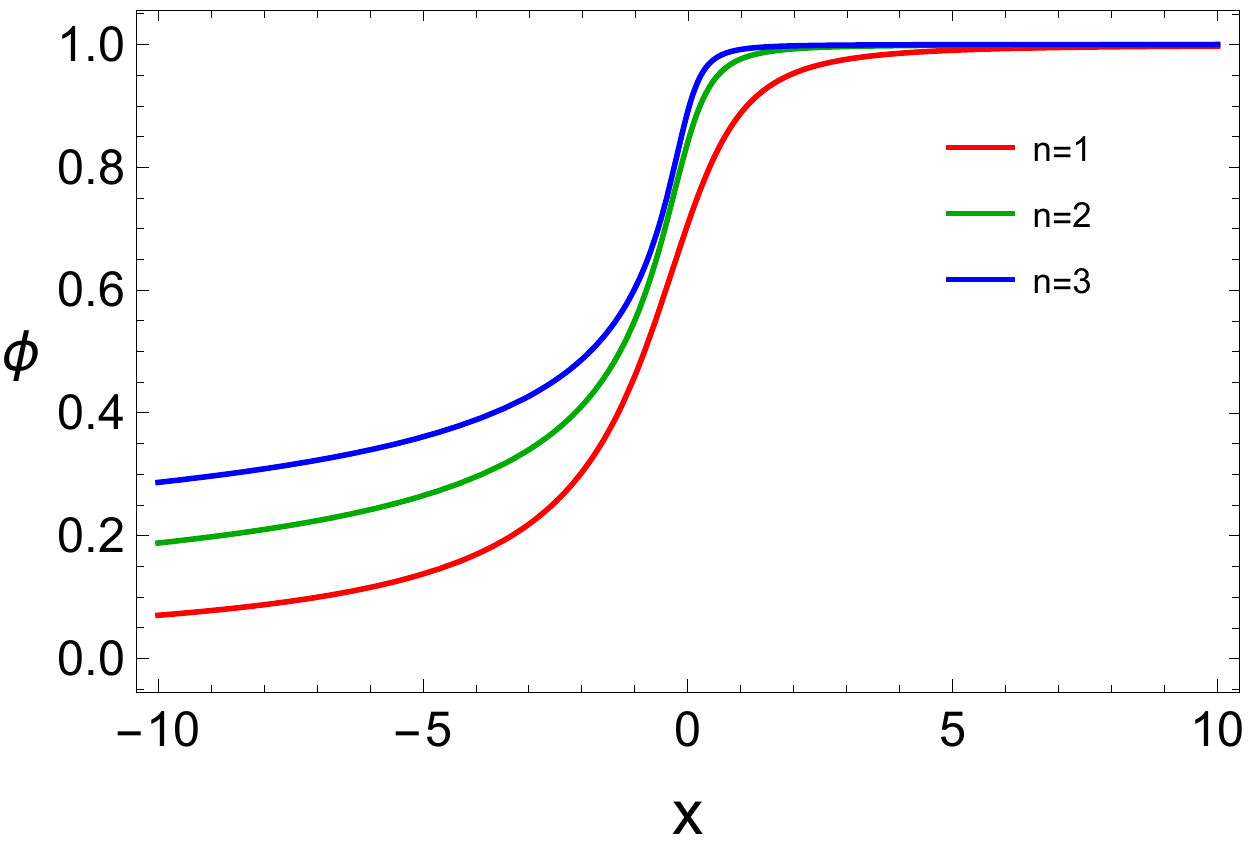}
\caption{Kink solution $\phi(x)$ given by Eq. (93) for $a=1$ and $\mu=\sqrt{2}n$. }
\end{figure} 

It is easy to calculate the kink mass in this case. We find
\be\label{3.6}
M_K = \int_{0}^{a} d\phi\, \sqrt{2V(\phi)}  
= \frac{3 \lambda \sqrt{\pi} a^{4n+2}}{4\sqrt{2} n} 
\frac{\Gamma[(n+2)/2n]}{\Gamma[(3n+1)/n]}\,.
\ee

Using the explicit kink solution (\ref{3.4}) it is straightforward to 
calculate the zero mode and we find
\be\label{3.7}
\psi_{0,K} \propto \frac{d\phi_{K}(x)}{dx} \propto 
 \frac{1}{\left[1+\frac{\mu x}{\sqrt{\mu^2 x^2+4}}\right]^{1-1/2n} (1+\mu^2 x^2)^{3/2}}\,.
\ee
As expected, this zero mode vanishes as $x\rightarrow\pm\infty$, i.e. as 
$\phi \rightarrow 0, a$. 

Finally the kink stability potential $V_{K}(x)$ which appears in the 
Schr\"odinger-like equation (\ref{9s}) 
is easily calculated
\bea\label{3.8}
V_{K}(x) &=& \frac{d^2V(\phi)}{d\phi^2}\Bigg|_{\phi = \phi_{K}(x)} 
= \frac{\lambda^2 a^{8n}}{16} \phi_{K}^{2n}(x)(a^{2n}-\phi_{K}^{2n}) 
\bigg [(2n+1)(2n+2)(a^{2n}-\phi_{K}^{2n}(x))^2 \nonumber \\
&-& 18n(2n+1)\phi_{K}^{2n}(a^{2n}-\phi_{K}^{2n}(x))
+24 n^2 \phi_{K}^{4n}(x) \bigg ]\,.
\eea
As expected, this kink potential $V_{K}(x)$ vanishes as 
$x \rightarrow \pm \infty$
thereby confirming that indeed in this case there is no gap between the
zero mode and the beginning of the continuum. Further, one finds that
$V_{K}(x = 0) = -\frac{(4n^2+6n-1)}{8}$.   

\subsection{Kink-Kink and Kink-Antikink Force}

Using the recent Manton formalism \cite{Manton19} it is straightforward to predict the
acceleration $a$ and hence force $F$ in both the K-K and AK-K case and hence the
ratio of the two forces. We predict that for the model characterized by the
potential (69) (with $a = 1$ and $\sqrt{2} \lambda = 1$) 
\be\label{3.9}
a_{K-K} = \bigg [\frac{\Gamma(\frac{1}{2[2n+1]}) \Gamma(\frac{n}{2[n+1]})}
{2\sqrt{\pi} (n+1)} \bigg ]^{2(2n+1)/n} \frac{4n\Gamma(\frac{3n+1}{n})}
{3\sqrt{\pi}\Gamma(\frac{n+2}{2n})} \frac{1}{d^{2(2n+1)/n}}\,,
\ee
\be\label{3.10}
a_{AK-K} = \bigg [-\frac{\sqrt{\pi} \Gamma(\frac{n}{2[2n+1]})}
{\Gamma(-\frac{1}{2[n+1]})} \bigg ]^{2(2n+1)/n} \frac{4n\Gamma(\frac{3n+1}{n})}
{3\sqrt{\pi} \Gamma(\frac{n+2}{2n})} \frac{1}{d^{2(2n+1)/n}}\,,
\ee
where $d$ is the K-K or AK-K separation distance, and hence the ratio of the 
two forces is 
\be\label{3.11}
\frac{F_{AK-K}}{F_{K-K}} = -\left[\sin(\frac{\pi n}{2n+1}) \right]^{2(2n+1)}\,,
~~n = 1, 2, 3,... \,.
\ee

It may be noted that for the model characterized by Eq. (\ref{3.1}), the kink 
tail around $\phi = 1$ has very different behavior compared to that around
$\phi = 0$ and 
as a result the kink-antikink acceleration and hence the corresponding force has
very different behavior. Generalizing the new Manton formalism \cite{Manton19} for the 
fractional values of $n$, we predict that in this model the kink-antikink 
acceleration is given by 
\be\label{3.12}
a_{K-AK} = \bigg [-\frac{\sqrt{\pi} \Gamma(\frac{1}{6})}
{\Gamma(-\frac{2}{3})} \bigg ]^{6} \frac{2\sqrt{2} n\Gamma(\frac{3n+1}{n})}
{3\sqrt{\pi} \Gamma(\frac{n+2}{2n})} \left(\frac{1}{d^{6}}\right) \,. 
\ee

\section{Explicit Kink Solutions with Power Law Tail for Another One-Parameter 
Family of Potentials} 
We now show that one can obtain exact kink solutions with power law tail in
yet another one-parameter family of higher order field theories characterized
by the potential 
\be\label{4.1}
V(\phi) = \lambda^2 \left|(a^{2n} - \phi^{2n})\right|^{(2n+1)/n}\,.
\ee
Note that this potential has degenerate minima at $\pm a$ and hence one kink
solution from $-a$ to $a$ and the corresponding antikink solution from $a$ to 
$-a$. Note that for the kink as well as the antikink solutions, $-a\le \phi \le a$. 

It is worth pointing out that whereas the potential (\ref{4.1}) is continuous, its
derivative is discontinuous at $x=\pm a$. However, since for the kink as well as
for the antikink solutions $-a\le \phi \le a$, this discontinuity would not matter
as far as the kink and the antikink solutions are concerned.
 
In order to obtain the kink solution from $-a$ to $a$, we need to solve the
self-dual equation 
\be\label{4.2}
\frac{d\phi}{dx} = \sqrt{2} \lambda \left|\phi^{2n}-a^{2n}\right|^{(2n+1)/2n}\,.
\ee
This is easily integrated yielding 
\be\label{4.3}
\frac{\phi}{(a^{2n}-\phi^{2n})^{1/2n}} = \mu x\,,
~~~\mu = \sqrt{2} \lambda a^{2n}\,.
\ee
Eq. (\ref{4.3}) is easily inverted yielding an explicit kink solution
\be\label{4.4}
\phi_{K}(x) = \frac{a \mu x}{[1+(\mu x)^{2n}]^{1/2n}}\,.
\ee
It is worth pointing out that for the special case of n = 1, the kink solution
(\ref{4.4}) has been obtained before \cite{Gomes12}. It is straightforward to 
find that
\be\label{4.5}
\lim_{x \rightarrow -\infty} \phi_{K}(x) = -a + \frac{a}{2n(\mu x)^{2n}}\,,
~~\lim_{x \rightarrow \infty} \phi_{K}(x) = a - \frac{a}{2n(\mu x)^{2n}}\,.
\ee
It is also straightforward to observe that 
\be\label{4.6}
M_K = \int_{-a}^{a} d\phi\, \sqrt{2V(\phi)}  
`= \frac{\sqrt{2} \lambda a^{2n+2}}{n} 
\frac{\Gamma[(4n+1)/2n] \Gamma[1/2n]}{\Gamma[(2n+1)/n]}\,.
\ee
Using the explicit kink solution (\ref{4.4}) it is easy to calculate the zero mode 
and we find 
\be\label{4.7}
\psi_{0,K} \propto \frac{d\phi_{K}(x)}{dx} \propto 
[1+(\mu x)^{2n}]^{-(2n+1)/2n}\,.
\ee
It is easily checked that, as expected the zero mode eigenfunction vanishes
as $x\rightarrow\pm\infty$, i.e. as $\phi \rightarrow\pm a$. 

Finally the kink stability potential $V_K(x)$ which appears in the Schr\"odinger-like 
Eq.~(\ref{9s}) is easily calculated using
\bea\label{4.8}
V_{K}(x) = \frac{d^2V(\phi)}{d\phi^2}\Bigg|_{\phi = \phi_{K}(x)} 
= 2(n+1)\lambda^2 a^{4n} \frac{(\mu x)^{4n-2}[4n-\frac{2n-1}{(\mu x)^{2n}}]}
{[1+(\mu x)^{2n}]^2}\,.
\eea
As expected, this kink potential vanishes as $x\rightarrow\pm\infty$ thereby 
confirming that indeed in this case there is no gap between the zero mode and 
the beginning of the continuum. 

\subsection{KK and K-AK Forces} 
Using the recent Manton formalism it is straight forward to predict the
acceleration a and hence force F in the AK-k case. However, it is not clear
if the recent Manton formalism \cite{Manton19} which was developed for 
calculating 
the force between the kink and the mirror kink also applies for the single kink
case. However since it applies for the AK-K case, we have an hunch that it
can also be extended to the calculation of the single K-K acceleration and
hence the force and hence the ratio of the AK-K vs K-K forces. Assuming
this, we predict that for this model characterized by the potential (81) (with
$a=1$ and $\sqrt{2}\lambda=1$) 
\be\label{4.9}
a_{K-K} = \bigg [\frac{n \Gamma(\frac{1}{2[2n+1]}) \Gamma(\frac{n}{2n+1})}
{\sqrt{\pi} (2n+1)} \bigg ]^{2(2n+1} \frac{n\Gamma(\frac{2n+1}{n})}
{2\Gamma(\frac{4n+1}{2n}) \Gamma(\frac{1}{2n})} \frac{1}{d^{2(2n+1)}}\,,
\ee
\be\label{4.10}
a_{K-AK} = \bigg [-\frac{\sqrt{\pi} \Gamma(\frac{1}{2[2n+1]})}
{\Gamma(-\frac{n}{2n+1})} \bigg ]^{2(2n+1} \frac{n\Gamma(\frac{2n+1}{n})}
{2\Gamma(\frac{4n+1}{2n}) \Gamma(\frac{1}{2n})} \frac{1}{d^{2(2n+1)}}\,,
\ee
where $d$ is the K-K or AK-K separation distance, and hence the ratio of the
two forces is 
\be\label{4.11}
\frac{F_{K-AK}}{F_{K-K}} = -\left[\sin(\frac{\pi n}{2n+1}) \right]^{2(2n+1)}\,.
\ee

\section{Summary and Possible Open Problems} 

In this paper we have presented five one parameter family of models for
which explicit kink solutions can be obtained. In three of these models, the kink tail 
has an exponential fall-off and using the standard Manton formalism \cite{Manton79} the
K-K and K-AK forces and hence their ratio is easily calculated. On the other
hand, in the other two models, the kink tail has a power-law fall off. To the best of 
our knowledge, these two models are the first two known higher order field theory 
examples for which explicit kink solutions with power-law tail can be obtained. Using
the recent Manton formalism \cite{Manton19} the K-K and AK-K forces and hence 
their ratio is easily calculated in the model discussed in Sec. V. However, it is
not obvious if the new Manton formalism is also applicable to calculate the
K-K force in the model of Sec. VI since in this case there is only a single
kink and there is no mirror kink. Note, however, that the AK-K force can
certainly be calculated in this case. We have assumed in Sec. VI that one
can extrapolate the new Manton formalism for this case too and using it we
have provided a prediction for the K-K acceleration and hence the corresponding 
K-K force as well as the ratio of the AK-K vs K-K force. It would be desirable 
if one could check whether this is indeed true or not.

Another open problem is to enquire if there exist higher order field theory
models which have super-exponential \cite{PKS19, Flores02}, super-super-exponential 
\cite{KS20PS} or power-tower \cite{KS20JPA} type of tails. One criticism against the 
models presented in Sec. V and VI is that these models are discontinuous at $x=\pm a$. 
While this should not matter as far as the calculation of the kink solutions in these models
is concerned since for all of them $-a\le x\le a$, it is certainly desirable to obtain explicit
kink solutions with power-law tail in models where no such criticism can be made. Yet 
another open problem is to understand why for kinks with power-law tail the ratio of the 
AK-K to K-K forces progressively decreases as we steadily go to higher order field theory 
models. This is unlike the kinks with exponential tails for which the magnitude of these 
two forces is always equal.

\vskip 0.1 in
\noindent{\bf Acknowledgment:}  A.K. is grateful to Indian National Science Academy (INSA) 
for the award of INSA Senior Scientist position at Savitribai Phule Pune University. 
The work at Los Alamos National Laboratory
was carried out under the auspices of the U.S. DOE and
NNSA under Contract No. DEAC52-06NA25396. 

\section{Appendix A: Additional Explicit Kink Solutions With Exponential Tail 
in Higher Order Field Theories} 

As mentioned at the end of Sec. 2, we now show that one can obtain exact kink 
solutions 
in a one-parameter family of higher order field theories characterized by the 
potential
\be\label{a1}
V(\phi) = \lambda^2 \phi^2 (a^{2n}-\phi^{2n})^2 (b^{2n}-\phi^{2n})^2\,,~~~
b > a\,,
\ee
in case $b^{2n} = 4 a^{2n}$. Note that the potential (\ref{a1}) has 5 
degenerate minima at $\phi = 0, \pm a, \pm b$
and hence two kink and two mirror kink solutions and corresponding four
antikink solutions. 

\subsection{\bf Kink Solution From $0$ to $a$ in case $b^{2n} = 4 a^{2n}$}

In this case the implicit kink solution takes the form
\be\label{a2}
e^{\mu f x} = \frac{\phi^{6n} (4a^{2n}-\phi^{2n})}{(\phi^{2n}-a^{n})^4}\,,
\ee
where $f$ is an arbitrary positive number satisfying $a^{2n} f = 1$ while 
$\mu$ now has a simple form
\be\label{a3}
\mu = \frac{24\sqrt{2} n \lambda}{f^3}\,.
\ee
Eq. (\ref{a2}) is easily inverted yielding a quartic equation  
\be\label{a4}
u^4 -\frac{6 u^2}{(1+e^{\mu f x})}+\frac{8 u}{(1+e^{\mu f x})}
-\frac{3}{(1+e^{\mu f x}+1)} = 0\,,
\ee
where $u = (1- \frac{\phi}{a})^{2n}$. Thus the explicit kink solution is 
given by
\bea\label{a5}
u = 1 - \Big(\frac{\phi}{a}\Big)^{2n} = 1 \mp \sqrt{2m}  
\mp \sqrt{\frac{12}{(1+e^{f \mu x})} 
- 2m \pm \frac{8}{\sqrt{m} (1+e^{\mu f x})}}\,,
\eea
where $m$ has the form 
\bea\label{a6}
m =\bigg [\frac{e^{\mu f x}[\sqrt{(1+e^{\mu f x})}+1}
{(1+e^{\mu f x})^{3}} \bigg ]^{1/3}  
+\bigg [\frac{e^{\mu f x}[\sqrt{(1+e^{\mu f x})}-1}
{(1+e^{\mu f x})^{3}} \bigg ]^{1/3} +\frac{2}{(1+e^{\mu f x})}\,.
\eea  

It is straightforward to check that
\be\label{a7}
\lim_{x \rightarrow -\infty} \phi(x) = \frac{a}{2^{1/3n}} e^{\mu f x/6n}\,,~~
\lim_{x \rightarrow \infty} \phi(x) = a -\frac{a 3^{1/4}}{2n} e^{-\mu f x /4}\,.
\ee

\subsection{Kink Solution From $a$ to $b$ in case $b^{2n} = 4 a^{2n}$}
 
In this case the implicit kink solution takes the form
\be\label{a8}
e^{-\mu f x} = \frac{\phi^{6n} (4a^{2n}-\phi^{2n})}{(\phi^{2n}-a^{n})^4}\,,
\ee
where $f$ is an arbitrary positive number satisfying $a^{2n} f = 1$ while 
$\mu$ now has a simple form as given by Eq. (\ref{a3}).

Eq. (\ref{a8}) is easily inverted yielding a quartic equation  
\be\label{a9}
u^4 -\frac{6 u^2}{(1+e^{-\mu f x})}-\frac{8 u}{(1+e^{-\mu f x})}
-\frac{3}{(1+e^{-\mu f x}+1)} = 0\,,
\ee
where $u = (\frac{\phi}{a})^{2n}-1$. Thus the explicit kink solution is 
given by
\bea\label{a10}
u = \Big(\frac{\phi}{a}\Big)^{2n}-1 = 1 \pm \sqrt{2m}  
\pm \sqrt{\frac{12}{(1+e^{-f \mu x})} 
- 2m \pm \frac{8}{\sqrt{m} (1+e^{-\mu f x})}}\,,
\eea
where $m$ has the form 
\bea\label{a11}
m =\bigg [\frac{e^{-\mu f x}[\sqrt{(1+e^{-\mu f x})}+1]}
{(1+e^{-\mu f x})^{3}} \bigg ]^{1/3}  
+\bigg [\frac{e^{-\mu f x}[\sqrt{(1+e^{-\mu f x})}+1]}
{(1+e^{-\mu f x})^{3}} \bigg ]^{1/3} +\frac{2}{(1+e^{-\mu f x})}\,.
\eea  

It is straightforward to check that
\be\label{a12}
\lim_{x \rightarrow -\infty} \phi(x) = a+\frac{a 3^{1/4}}{2n} e^{\mu f x/4}\,,~~
\lim_{x \rightarrow \infty} \phi(x) = 4^{1/2n}a 
-\frac{a 3^{4} 2^{1/n}}{2^{8}} e^{-\mu f x}\,.
\ee

\section{Appendix B: Analytical Kink Solutions with a Power-Law Tail in Yet 
Another One-Parameter Family of Higher Order Field Theories}

We show that one can obtain exact kink solutions with a power-law tail
in a one-parameter family of higher order field theories characterized by
the potential
\be\label{5.1}
V(\phi) = \lambda^2 \phi^{2(3n+1)} \left|(a^{2n} - \phi^{2n})\right|^{3}\,.
\ee
Note that this potential has degenerate minima at $ \phi = 0, \pm a$ and 
hence one
kink solution from $0$ to $a$ and the corresponding mirror kink solution
from $-a$ to $0$ and the corresponding two antikinks. Note that for both 
the kink solutions $-a \le \phi \le a$
and hence even though the potential is discontinuous at $\phi = \pm a$
that would not matter as far as the kink and corresponding antikink 
solutions are concerned.

\subsection{Explicit Kink Solution}

In order to obtain the kink solution, we need to solve the self-dual
equation
\be\label{5.2}
\frac{d\phi}{dx} = \sqrt{2} \lambda \phi^{3n+1} \left|\phi^{2n}-a^{2n}\right|^{3/2}\,.
\ee
This is easily integrated yielding
\be\label{5.3}
\frac{(8\phi^{4n} -4 \phi^{4n} -a^{4n})^2}{9 \phi^{6n} (a^{2n}-\phi^{2n})} 
= \mu^2 x^2\,,
~~~\mu = \sqrt{2}n \lambda a^{6n}\,.
\ee
Eq. (\ref{5.3}) is easily inverted yielding a quartic equation  
\be\label{5.3a}
u^4 +a u^2 +b u + d = 0\,,
\ee
where
\be\label{5.4}
a = -\frac{3}{8}\,,~~~b = -\frac{(192+99\mu^2 x^2)}{64 (64+9 \mu^2 x^2)}\,,
~~~d = \frac{(576- 27 \mu^2 x^2)}{256(64+9\mu^2 x^2)}\,,
\ee
while $u = (\frac{\phi}{a})^{2n} -1/4$. Thus the explicit kink solution is 
given by
\be\label{5.5}
u = \left(\frac{\phi}{a}\right)^{2n} - 1/4 = \pm \sqrt{2m} 
\pm \sqrt{-(2a+2m \pm \frac{2b}{\sqrt{2m}})}- a/3\,, 
\ee
where $m$  satisfies the cubic equation
\be\label{5.6}
8m^3+8a m^2 +(2a^2 -8d) m - b^2 = 0\,.
\ee
It is not difficult to check that in this case $m$ is given by 
\be\label{5.7}
m =\bigg [-\frac{b}{2}+\sqrt{\frac{b^2}{4}+\frac{a^3}{27}} \bigg ]^{1/3}
+\bigg [-\frac{b}{2}-\sqrt{\frac{b^2}{4}+\frac{a^3}{27}} \bigg ]^{1/3}\,.
\ee

It is straightforward to check that
\be\label{5.8}
\lim_{x \rightarrow -\infty} \phi_{K}(x) =  \frac{a}{(9\mu^2 x^2)^{1/6n}}\,,
~~~\lim_{x \rightarrow \infty} \phi_{K}(x) = a - \frac{a}{2n\mu^2 x^2}\,.
\ee

It is easy to calculate the kink mass in this case. We find
\be\label{5.9}
M_K = \int_{0}^{a} d\phi\, \sqrt{2V(\phi)}  
= \frac{3 \lambda \sqrt{\pi} a^{6n+2}}{4\sqrt{2}} 
\frac{\Gamma[(3n+2)/2]}{\Gamma[(3n+7)/2]}\,.
\ee

Finally, as expected in this case too there is no gap between the zero mode
and the beginning of the continuum. 

\subsection{\bf Kink-Kink and Kink-Antikink Force}

Using the recent Manton formalism it is straightforward to predict the acceleration $a$ 
and hence the force $F$ in both the K-K and AK-K cases and thus the
ratio of the two forces. We predict that for the model characterized by the
potential (\ref{3.1}) (with $a = 1$ and $\sqrt{2} \lambda = 1$) 
\be\label{5.10}
a_{K-K} = \bigg [\frac{\Gamma(\frac{1}{2[3n+1]}) \Gamma(\frac{3n}{2[3n+1]})}
{2\sqrt{\pi} (3n+1)} \bigg ]^{2(3n+1)/3n} \frac{4\Gamma(\frac{3n+7}{2})}
{3\sqrt{\pi}\Gamma(\frac{3n+2}{2})} \frac{1}{d^{2(3n+1)/3n}}\,,
\ee
\be\label{5.11}
a_{AK-K} = \bigg [-\frac{\sqrt{\pi} \Gamma(\frac{3n}{2[3n+1]})}
{\Gamma(-\frac{1}{2[3n+1]})} \bigg ]^{2(3n+1)/3n} \frac{4\Gamma(\frac{3n+7}{2})}
{3\sqrt{\pi} \Gamma(\frac{3n+2}{2})} \frac{1}{d^{2(3n+1)/3n}}\,,
\ee
where $d$ is the K-K or AK-K separation distance, and hence the ratio of the 
two forces is 
\be\label{5.12}
\frac{F_{AK-K}}{F_{K-K}} = - \left[\sin\left(\frac{\pi}{2[3n+1]}\right) 
\right ]^{2(3n+1)/3n}\,,
~~n = 1, 2, 3,... \,.
\ee

It may be noted that for the model characterized by Eq. (\ref{5.1}), the kink 
tail around $\phi = 1$ has very different behavior compared to that around
$\phi = 0$ and as a result the kink-antikink acceleration and hence the 
corresponding force have very different behavior. Generalizing the new 
Manton formalism \cite{Manton19}, we predict that in this model the kink-antikink 
acceleration is given by 
\be\label{5.13}
a_{K-AK} = \bigg [-\frac{\sqrt{\pi} \Gamma(\frac{1}{6})}
{\Gamma(-\frac{1}{3})} \bigg ]^{6} \frac{1}{8n^3} \frac{4\Gamma(\frac{3n+7}{2})}
{3\sqrt{\pi} \Gamma(\frac{3n+2}{2})} \left(\frac{1}{d^{6}}\right) \,.
\ee

\section{Appendix C: Known Explicit Solutions in Higher Order Field Theory Models} 

For the sake of completeness, we list here the previously known solutions from 
\cite{Chapter, KS19, GaniPRD20}. To the best of our knowledge, so far only three 
explicit kink solutions are known for higher order (polynomial) field theories, i.e. for 
$\phi^8, \phi^{10}, \phi^{12}$ and even higher order field theories.  Two of these 
are with exponential tails while one has a power-law tail. In particular, while 
explicit kink solution with a power-law tail is known for a $\phi^{10}$ model
\cite{KS19}, explicit kink solutions with an exponential tail are known for a 
(i) $\phi^{4n+2}-\phi^{2n+2}-\phi^2$ ($n = 1, 2, 3,...$) model \cite{KS19} and (ii) for 
a $\phi^{8}$ model \cite{GaniPRD20}. 

\subsection{\bf $\phi^{10}$ Model with a Power-Law Tail}

It has been shown \cite{KS19} that for the kink potential
\be\label{b1}
V(\phi) = \lambda^2 \left|(a^2-\phi^2)\right|^{5}\,,
\ee
the explicit kink solution from $-a$ to +$a$ is given by
\bea\label{b2}
1-(\frac{\phi_{K}}{a})^2 = \bigg [\frac{\sqrt{4+\mu^2 x^2}+\mu x}
{2(4+\mu^2 x^2)^{3/2}} \bigg ]^{1/3} 
+ \bigg [\frac{\sqrt{4+\mu^2 x^2}-\mu x}
{2(4+\mu^2 x^2)^{3/2}} \bigg ]^{1/3}\,.
\eea
The details can be found in the reference mentioned above.

\subsection{\bf One-Parameter Family of Potentials with an Exponential Tail}

It has been shown \cite{KS19} that for the one-parameter family of potentials
\be\label{b3}
V(\phi) = \frac{\lambda^2}{2} \phi^2 (a^{2n}-\phi^{2n})^2\,, 
~~n = 1, 2, 3,... \,,
\ee
which has a kink solution from $0$ to $a$ and a mirror kink from $-a$ to $0$
and the corresponding two antikinks, the kink solution from $0$ to $a$ is 
given by
\be\label{b4}
\phi_{K} = a \left[\frac{1+\tanh(\beta x)}{2}\right]^{1/2n}\,, 
\ee
where $\beta = n \lambda a^{2n}$. 
The details can be found in the reference mentioned above.

\subsection{\bf $\phi^8$ Model with an Exponential Tail}

It has been shown \cite{GaniPRD20} that for the $\phi^{8}$ model
\be\label{b5}
V(\phi) = (1/2) (a^2-\phi^2)^2 (b^2-\phi^2)^2\,,~~~0 < a < b\,,
\ee
which has kink solutions from $a$ to $b$, mirror kinks from $-b$ to $-a$ and
a kink solution from $-a$ to +$a$. In case $b = 1, a = 1/2$, the three kink
solutions and the corresponding antikink solutions are given by
\be\label{b6}
\phi_{K} = \cos \bigg [\frac{1}{3}\cos^{-1}\Big(\tanh \frac{3x}{4}\Big) +\frac{\pi m}{3} 
\bigg ]\,,
\ee
depending on the value of $m = 0, 1, 2, 3, 4, 5$. On the other hand, in case
$b = 1, a = 1/4$ the three kink and three antikink solutions are given by
the solution of a quartic equation. The details can be found in the reference
mentioned above.

\end{document}